\documentclass[journal]{IEEEtran}
\usepackage{amsfonts}
\IEEEoverridecommandlockouts

\ifCLASSINFOpdf
\else
\fi
\usepackage{subfigure}
\usepackage{algorithm}
\usepackage{algorithmic}
\usepackage{psfig}
\usepackage{url}
\usepackage{epstopdf}
\usepackage{multirow}
\usepackage{cite}
\usepackage{stfloats}
\usepackage{color}
\usepackage{epsfig}
\usepackage{graphicx}
\usepackage{psfig}
\usepackage{epsf}
\usepackage{slashbox}
\usepackage{float}
\usepackage{amssymb }
\usepackage[cmex10]{amsmath}
\usepackage{algorithm} 
\usepackage{algorithmic}
\usepackage{booktabs}
\usepackage{amsmath,bm}
\usepackage{subfigure}
\begin{document}

\title{On the Outage Performance of Ambient Backscatter Communications}
\author{Yinghui Ye, Liqin Shi, Xiaoli Chu,~\IEEEmembership{Senior Member,~IEEE}, and Guangyue Lu
\thanks{Copyright (c) 20xx IEEE. Personal use of this material is permitted. However, permission to use this material for any other purposes must be obtained from the IEEE by sending a request to pubs-permissions@ieee.org.}
\thanks{Yinghui Ye and Liqin Shi  are with the Shaanxi
Key Laboratory of Information Communication Network and Security, Xi'an
University of Posts \& Telecommunications, China, and are also with the School of Telecommunication Engineering, Xidian University,  China. (e-mail: connectyyh@126.com, liqinshi@hotmail.com) }
\thanks{Xiaoli Chu is with the Department of Electronic and Electrical Engineering, The University of Sheffield, U.K. (e-mail: x.chu@sheffield.ac.uk)}
\thanks{Guangyue Lu is with the Shaanxi Key Laboratory of Information Communication Network and Security, Xi'an University of Posts and Telecommunications (e-mail: tonylugy@163.com)}
\thanks{This work was supported by the  Natural Science Foundation of China (61801382) and the Science and Technology Innovation Team of Shaanxi Province for Broadband Wireless and Application (2017KCT-30-02).}
}
\maketitle
\begin{abstract}
  A{\color{black}mbient backscatter communications (AmBackComs) have been recognized as a spectrum- and energy-efficient technology for Internet of Things},
 as it allows  passive backscatter devices (BDs) to modulate their information into the legacy signals, e.g., cellular signals,  and reflect them to their associated receivers while harvesting energy from the legacy signals to power their circuit operation. {\color{black} However, the co-channel interference between   the backscatter link and the legacy link and the non-linear behavior of energy harvesters at the BDs have largely been ignored in  the performance analysis of AmBackComs.  Taking these two aspects, this paper provides a comprehensive outage performance analysis for an AmBackCom system with multiple backscatter links}, where  one of the backscatter  links is opportunistically selected to leverage the legacy signals transmitted in a given resource block. For any selected backscatter link, we propose an adaptive reflection coefficient (RC), which is adapted to the non-linear energy harvesting (EH) model and the location of the selected backscatter link, to minimize the outage probability of the backscatter link.
 In order to study the impact of co-channel interference on  both  backscatter and  legacy links,  for a selected backscatter link,
we derive the outage probabilities for the legacy link and the backscatter link.
Furthermore, we study the best and worst outage performances for the backscatter system where the selected backscatter link maximizes or minimizes the  signal-to-interference-plus noise ratio (SINR)  at the backscatter receiver. We also study the best and worst outage performances for the legacy link where the selected backscatter link results in the lowest and highest co-channel interference to the legacy receiver, respectively. Computer simulations validate our analytical  results, and reveal the impacts of the co-channel interference and  the EH model on the AmBackCom performance. In particular, the co-channel interference leads to the  outage saturation phenomenon in AmBackComs, and the conventional linear EH model results in an over-estimated outage performance for the backscatter link.
\end{abstract}
\IEEEpeerreviewmaketitle
\section{Introduction}
\IEEEPARstart{I}n the era of Internet of Things (IoT), a great number of  devices will be deployed to monitor, sense, and generate enormous  data, in order to support different applications, e.g., smart  city, and intelligent agriculture. It is predicted by Ericsson that  about 22.3 billion IoT  devices will be deployed worldwide by   2024 \cite{2222}. The  development of massive IoT devices is facing different challenges caused by the limited battery capacity of IoT deveices  and the limited spectrum resource.
In this context, {\color{black} ambient backscatter  communication (AmBackCom) has been considered as a potential solution to address these two challenges  \cite{8368232}, \cite{6G}.
In AmBackComs, the backscatter device (BD) modulates its message on the received legacy signals, e.g., cellular or WiFi signals, and reflects  the modulated signals to its associated receiver, while harvesting energy from the legacy signals for covering the circuit energy consumption \cite{6742719,8730429}. This is different from the simultaneous
wireless information and power transfer (SWIPT), where the transmitter generates radio frequency (RF) signals itself for conveying energy and information to the receiver simultaneously. Thus, the BD does not need  active components, e.g., oscillators, analog-to-digital/digital-to-analog converters, resulting in a much less
energy consumption as compared with a SWIPT transmitter \cite{8368232,kellogg2014wi}.
 Meanwhile, the backscatter link shares  the same spectrum resource with the legacy link, i.e.,  no extra spectrum resource is required for AmBackComs.}

Due to the spectrum sharing between the legacy link and the backscatter link,  there will be co-channel interference between them. Symbol detectors  have been proposed for THE backscatter receiver (BR) to  suppress the co-channel interference caused by the legacy transmitter (LT).
{\color{black} In \cite{7551180}, for the
differential on-off modulation,  a maximum a posteriori (MAP)
based detector was proposed by exploiting  the difference between two
consecutive signal powers. In \cite{8007328}, a semi-coherent detection was developed based on the likelihood
ratio test. Exploiting the advantages of multiple antennas,
the authors of \cite{8865663} proposed a maximum-eigenvalue detector
and demonstrated its superior symbol detection performance
by comparing with the energy-based detector. In \cite{8103807},  the ambient waveform
and the symbol detector were jointly designed for orthogonal frequency division multiplexing (OFDM) systems. The above
works mainly focus on the detectors design and characterizing
the achievable bit error rate  for backscatter links, but have not considered
the outage performance. }

In addition to   \cite{8007328,8865663,8103807}, there are a considerable number of studies on the  design of  resource allocation schemes for AmbackComs.
{\color{black}In \cite{7937935, 8253544, 8340034,8632710,7981380,8327597, 7996410, 8834802}, the authors proposed  harvest-then-transmit (HTT) enabled AmBackComs, and designed  resource allocation schemes to satisfy various optimization goals.}
In these works, they mainly focused on balancing the time allocation  between the backscatter mode and the HTT mode in order to maximize the achievable throughput of  backscatter links based on  a fixed transmission rate of backscatter links, i.e., the reflection coefficient (RC) of  BDs and the transmit power of the LT are fixed. By assuming  that both the RC of BDs and the transmit power of the LT can be adjusted, the resource allocation scheme was  designed in the coexistence of  cognitive  networks and AmbackComs \cite{8424210}, where the  sum throughput of the backscatter link is maximized.   AmbackCom was  combined with  full-duplex transmissions in  \cite{8476159}, where a full-duplex access point  simultaneously transmits signals to its legacy receiver (LR) and  receives the backscattered signals from BDs, and the fairness among the backscatter links in terms of the achievable throughput is guaranteed.

{\color{black} The performance analysis in terms of ergodic capacity and outage probability has also been investigated in AmBackComs. In \cite{8345348}, the ergodic capacity of the backscatter link with a two-state modulation was derived in the case of real or complex RF signals. In \cite{7820135}, the authors analyzed the ergodic capacity in the coexistence of AmBackComs and legacy systems that employ an OFDM scheme. In \cite{8807353}, the authors analyzed the ergodic rate for legacy and backscatter links, where the BD and the BR are co-located. In \cite{8922800},
the author proposed an opportunistic AmBackCom-assisted decode-and-forward relay network, and derived a closed-form expression of the achievable ergodic capacity. The
outage probability of a backscatter link was analyzed in  \cite{8468064}, where multiple antennas and maximum ratio combing were employed at the BR, but the outage performance of the legacy
link was not studied. Using tools from stochastic geometry, the outage probabilities and achievable rates for the legacy and backscatter links were investigated \cite{8377363}. The authors in \cite{8370749} derived
the outage probability of the backscatter link for a two-state modulation at the BD. The authors of \cite{8360017} considered the backscatter links as the secondary users in a cognitive relay network, and derived the outage probabilities of both the primary and backscatter links. In \cite{8170328}, the authors proposed a novel hybrid device-to-device (D2D) transmitter that alternately operates in the backscatter mode and the HTT mode, and analyzed the outage probability and the average outage capacity. However, there are limitations in the above existing works, which are listed below.}

\begin{table*}
\centering
\begin{tabular}{|l|l|}
\multicolumn{2}{c}{\textbf{{\color{black}Table I Definitions of Notations}}}\\
\hline
\color{black} Notation  & \color{black} Meaning  \\
\hline

\color{black} $\alpha$ &  \color{black} Path loss exponent\\
\hline
\color{black} $\mathbb{P}(\cdot)$ & \color{black}  Probability operator \\
\hline
\color{black} $\mathbb{E}[\cdot ]$ & \color{black} Expectation operator \\
\hline
\color{black} $P_t$ &\color{black} Transmit power of the LT\\
\hline
\color{black}$\Lambda_{1k}$ & \color{black}Frequency dependent constant \\
\hline
\color{black}$\beta_k$ & \color{black} Reflection coefficient of the $k$-th BD\\
\hline
\color{black}$P_{c,k}$ & \color{black}Circuit power consumption  at the $k$-th BD\\
\hline
\color{black}$T$ & \color{black}A whole time slot of the legacy transmission \\
\hline
\color{black} $\mathcal{CN}(a,b)$ &\color{black}  Gaussian distribution with mean $a$ and variance $b$\\
\hline
\color{black}$\gamma^b_{\rm{th}}$ and $\gamma_{\rm{th}}$ &   \color{black}SINR threshold for the $k$-th backscatter link and the  legacy link \\
\hline
\color{black}$\mathcal{P}^b_{{\rm{out}},k}$ and $\mathcal{P}^l_{{\rm{out}},k}$ & \color{black} Outage probability for the $k$-th backscatter link and the  legacy link\\
\hline
\color{black}$h_p$ and $d_p$ & \color{black}Channel coefficient and  distance of the legacy link\\
\hline
\color{black}$g^p_{k}$ and $d^p_{k}$ & \color{black} Channel coefficient  and distance of the LT-to-the $k$-th  BR link $\left(k\in\{1,\cdot\cdot\cdot,K\}\right)$ \\
\hline
\color{black}$h_{1k}$ and $d_{1k}$&   \color{black}Channel coefficient and distance of the LT-to-the $k$-th BD link $\left(k\in\{1,\cdot\cdot\cdot,K\}\right)$\\
\hline
 \color{black}$h_{2k}$ and $d_{2k}$ &  \color{black} Channel coefficient and distance of the $k$-th backscatter link $\left(k\in\{1,\cdot\cdot\cdot,K\}\right)$\\
\hline
\color{black}$g^s_{k}$ and $d^s_{k}$ & \color{black} Channel coefficient and distance of the $k$-th BD-to-LR link $\left(k\in\{1,\cdot\cdot\cdot,K\}\right)$ \\
\hline
\end{tabular}
\end{table*}
\begin{itemize}
\item The studies in \cite{7820135,8468064,8922800,8360017,8807353} assumed that the BR was able to successfully decode the legacy signals and perform a perfect successive interference cancellation (SIC) to remove the interference brought by the LT. The authors of  \cite{8377363,8170328} ignored the interference from the LT to the backscatter links. In addition, the outage performance of legacy links has largely been ignored  \cite{8468064,8170328,8370749}. Therefore, there is  a lack of a comprehensive performance analysis for AmbackComs.
\item  {\color{black}In AmBackComs, an energy outage event at the backscatter link occurs when the harvested energy at the BD cannot meet  the circuit consumption. Hence, the performance analysis of AmBackComs requires  an accurate model for the harvested energy. In \cite{8345348} and \cite{8370749}, the energy outage event was not considered. The studies \cite{8922800,7937935, 8253544, 8340034,8632710,8476159,8424210,7981380, 7820135,7996410, 8327597,8468064,8377363,8360017,8807353,8170328} assume  a linear energy harvesting (EH) model at the BD, which fails to    characterize the inherent non-linearity of a practical energy harvester \cite{8060616,8355777,4494663}. The use of an inaccurate EH model will affect the accuracy of signal-to-interference-plus-noise ratios (SINRs) for both legacy and backscatter links. Thus, it is desirable to consider a non-linear EH model in the performance analysis for AmBackComs.}
\end{itemize}


In this paper, we take the above observations into account and analyze the outage probability of an AmBackCom network. In particular,
we consider an AmBackCom network with multiple backscatter links, where only one backscatter link is opportunistically selected to   backscatter the legacy signal transmitted in any given resource block, thus avoiding severe co-channel interference among backscatter links.


Our main contributions are summarized as follows.
\begin{itemize}
\item {\color{black}Considering a practical non-linear EH model, we propose an adaptive RC scheme to minimize the outage probability for any selected backscatter link and obtain a closed-form expression for the optimal RC, which provides a guideline for BDs to set the value of RC in a practical backscatter communication system. With the optimal RC and a selected backscatter link,  the outage probabilities for the backscatter link and the legacy link are derived. The outage-probability lower  bounds for both links  are  obtained as the transmit power of the LT approaches infinity. 
\item  
   We evaluate  the best and worst outage performances for the
backscatter system and the legacy transmission, respectively. Specifically, the selected backscatter link with the maximum (or minimum) SINR at the BR leads to the best (or worst) outage performance for the backscatter system, while the selected backscatter resulting in the lowest (or highest) co-channel interference to the LR causes the best (or worst) outage performance for the legacy system.
\item We examine the effects  of  co-channel interference and the EH model on the achievable outage probability of the considered AmBackCom network, and make the following new observations.  Firstly, the co-channel interference between  the backscatter link and the legacy link leads to an outage saturation phenomenon for both the backscatter and legacy links at a high transmit power of the LT. Secondly, the  passive BD that causes  the lowest co-channel interference to the LR can realize its own information transmission at the cost of a slightly degraded outage performance of the legacy link even for    low or medium transmit power of the  LT. Thirdly, the conventionally used  linear EH model leads to a greatly over-estimated outage performance for the backscatter link and a slightly underrated outage performance for the legacy link.}
\end{itemize}

\begin{figure*}
  \centering
  \includegraphics[width=0.94\textwidth]{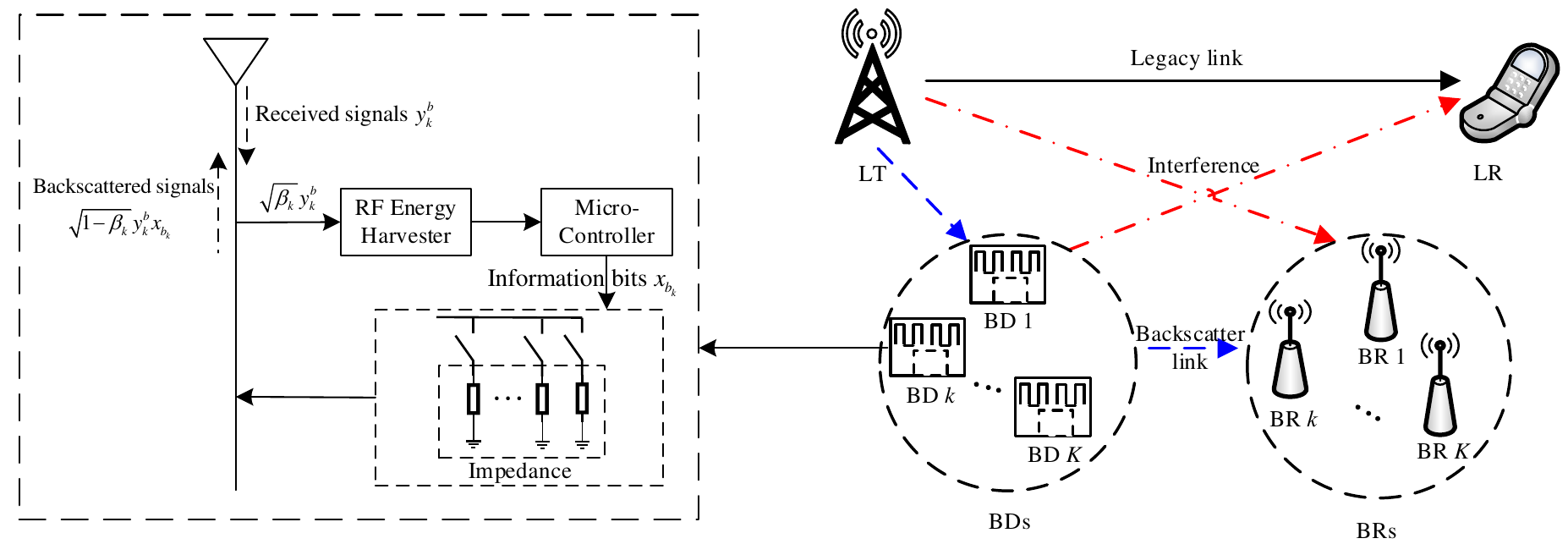}\\
  \caption{ \color{black}System model: the left subgraph subplot shows the structure of the $k$-th BD, and the right subgraph subplot illustrates the AmBackCom network.  }\label{fig0}
\end{figure*}

The remainder of this paper is organized as follows. The system model is provided in Section \uppercase\expandafter{\romannumeral 2}.
Sections \uppercase\expandafter{\romannumeral 3} and \uppercase\expandafter{\romannumeral 4}
analyze the outage performance of the backscatter system and the legacy transmission, respectively.
Numerical results are shown in Section \uppercase\expandafter{\romannumeral 5}, followed by conclusions in Section \uppercase\expandafter{\romannumeral 6}. {\color{black}Besides,
the main notations have been summarized in Table I, as shown at the top of the next page.}

\section{System Model}
In this work, we consider an AmBackCom network, which consists of one LT, one LR, and $K$ BDs and their receivers, as shown in Fig. 1. In this network,
the LT transmits its  signal to the LR  in a given resource
block, forming a legacy radio link, while one of  backscatter links is opportunistically selected to modulate its information into the LT transmitted signal and backscatter the modulated signal to its own receiver.
Assume that all channels are quasi-static and subject to path-loss and Rayleigh fading.
Let $h_p$, $g^p_{k}$, $h_{1k}$, $h_{2k}$ and $g^s_{k}$ $\left(k\in\{1,\cdot\cdot\cdot,K\}\right)$ denote the channel coefficients\footnote{ {\color{black}The channel gains $|g_k^p|^2$ and $|h_{1k}|^2|h_{2k}|^2$ can be estimated by using the channel estimation method proposed in \cite{8320359}. Please note that $|h_{1k}|^2$ can be obtained after estimating $|h_{2k}|^2$, and
$|h_{2k}|^2$ can be estimated  using  traditional channel estimation methods. Particularly,
the $k$-th BR is a traditional node  and can send a pilot signal to the $k$-th BD and then the $k$-th BD reflects the  pilot signal to the $k$-th BR. By performing least-square estimation, the $k$-th BR can obtain the product of the forward channel gain  and the backward channel gain. Due to the channel reciprocity, the forward channel gain equals the  backward channel gain and hence $|h_{2k}|^2$  is obtained by taking the square root of the product.}} of the legacy link, the LT-to-the $k$-th  BR link, the LT-to-the $k$-th BD link, the $k$-th backscatter link, and the $k$-th BD-to-LR link, respectively.
The corresponding distances are denoted by $d_p$, $d^p_{k}$, $d_{1k}$, $d_{2k}$ and $d^s_{k}$, respectively.

For the legacy system, the LT transmits its information $x_p$ with {\color{black}$\mathbb{E}[|x_p|^2]=1$} to the LR. Meanwhile, the received  signal at the  $k$-th BD is given by
\begin{align}\label{1}
y^b_{k}=\sqrt{P_t\Lambda_{1k}d_{1k}^{-\alpha}}h_{1k}x_p+n^b_k,
\end{align}
where $P_t$ is the transmit power of the LT; $\Lambda_{1k}$ denotes the frequency dependent constant \cite{8633928} for the LT-to-the $k$-th BD link; $\alpha$ is the path loss exponent and $n^b_k\sim\mathcal{CN}(0,\sigma^2)$ denotes the additive white Gaussian noise (AWGN)  at the $k$-th BD.

Following \cite{8424210},  the received signals at the $k$-th BD is divided into two parts through a RC $\beta_k$:  $\sqrt{\beta_k}y^b_{k}$  used for EH and the remaining one with $\sqrt{1-\beta_k}y^b_{k}$  for backscattering.
Here we consider a practical non-linear EH model proposed in \cite{8060616} to characterize the harvested power.
Then the total harvested energy at the $k$-th BD is given by
\begin{align}\label{4aa}
E_{k} = \frac{{{E_{\max ,k}}\left( {1 - \exp \left( { - {s_{1k}}{P_{r,k}} + {s_{1k}}{s_{0k}}} \right)} \right)}}{{1 + \exp \left( { - {s_{1k}}{P_{r,k}}{\rm{ + }}{s_{1k}}{s_{2k}}} \right)}}T,
\end{align}
where $P_{r,k}=\beta_kP_t\Lambda_{1k}d_{1k}^{-\alpha}|h_{1k}|^2$ is the input RF power of the harvester at the $k$-th BD; ${E_{\max,k}}$ is the maximum harvestable power when the circuit is saturated; $s_{0k}$
denotes the sensitivity threshold; $s_{1k}$ and $s_{2k}$ are fixed parameters determined by the resistance, capacitance, and diode turn-on voltage; $T$ is the whole time slot for the legacy transmission.

 Let $P_{c,k}$ denote the circuit power consumption  at the $k$-th BD.
 When $E_k\geq P_{c,k}T$, the $k$-th BD has enough energy for the circuit operation. Otherwise, the $k$-th BD can not backscatter information to its receiver. For the above two cases, we can write the received SINR at the BR and LR as follows.


(i) If $E_k\geq P_{c,k} T$, the $k$-th BD can modulate its own information $x_{b_k}$ with {\color{black}$\mathbb{E}[|x_{b_k}|^2]=1$} on the received signal $\sqrt{1-\beta_k}y^b_{k}$, and reflect the modulated signals. Then the received signal from the $k$-th BD to its receiver is
\begin{align}\label{3}\notag
y^r_{k}&=\sqrt {\eta_k \left( {1 - {\beta _k}} \right){P_t}{K_{1k}}{K_{2k}}} {h_{1k}}{h_{2k}}{x_p}{x_{{b_k}}}\\
& + \sqrt {{P_t}{K^p_{k}}} g_k^p{x_p} + {n^r_{k}},
\end{align}
where $\eta_k$ denotes the backscatter efficiency \cite{6742719} for the $k$-th BD, ${K_{1k}} = {\Lambda _{1k}}d_{1k}^{ - \alpha },{K_{2k}} = {\Lambda _{2k}}d_{2k}^{ - \alpha },K_k^p = \Lambda _k^p{\left( {d_k^p} \right)^{ - \alpha }}$ with the frequency dependent constant for the LT-to-the $k$-th BR link $\Lambda _k^p$, and ${n^r_{k}}\sim\mathcal{CN}(0,\sigma^2)$ is the AWGN at the $k$-th BR.
Accordingly,  the SINR for decoding $x_{b_k}$ at the BR can be written as
  \begin{align}
   {\gamma^b _k} = \frac{{\eta_k \left( {1 - {\beta _k}} \right){P_t}{K_{1k}}{K_{2k}}{{\left| {{h_{1k}}} \right|}^2}{{\left| {{h_{2k}}} \right|}^2}}}{{{P_t}{K_{k}^p}{{\left| {g_k^p} \right|}^2} + {\sigma ^2}}}.
  \end{align}

Since the backscattered signals cause interference to the LR, the received signal at the LR is given by
\begin{align}\notag\label{4}
y^{(1)}_k= &\sqrt {{\eta _k}\left( {1 -{\beta _k}} \right){P_t}{K_{1k}}K_k^s} {h_{1k}}g_k^s{x_p}{x_{{b_k}}} \\
&+ \sqrt {{P_t}{K_p}} {h_p}{x_p}+{n_p}
\end{align}
where ${K_p} = {\Lambda _p}d_p^{ - \alpha }$, $K_k^s = \Lambda _k^s{\left( {d_k^s} \right)^{ - \alpha }}$, ${\Lambda _p}$ and $\Lambda _k^s$ are the frequency dependent constants for the legacy link and the $k$-th BD-to-LR link, respectively; ${n_p}\sim\mathcal{CN}(0,\sigma^2)$ is the AWGN at the LR.
Thus the SINR at the LR is written as
\begin{align}
{\gamma ^{(1)}_k} = \frac{{{P_t}{K_p}{{\left| {{h_p}} \right|}^2}}}{{{\eta _k}\left( {1 - {\beta _k}} \right){P_t}{K_{1k}}K_k^s{{\left| {{h_{1k}}} \right|}^2}{{\left| {g_k^s} \right|}^2} + {\sigma ^2}}}.
\end{align}

(ii) If $E_k< P_{c,k} T$, the $k$-th BD keeps silent and the $k$-th BR link  is in outage.
In this case,  the received signal at the LR is written as
\begin{align}\label{4-1}
y^{(2)}_k=\sqrt {{P_t}{K_p}} {h_p}{x_p}+ {n_p}.
\end{align}
Then the signal-to-noise-ratio  (SNR) at the LR is calculated as
\begin{align}
{\gamma ^{(2)}_k} = \frac{{{P_t}{K_p}{{\left| {{h_p}} \right|}^2}}}{{\sigma ^2}}.
\end{align}

\section{Outage Analysis for AmBackComs}
In this section, we analyze the outage performance\footnote{\color{black}Reliable transmission and  the
performance evaluation are  crucially important in wireless communication systems. The outage probability has been recognized as one of important  performance metrics to
evaluate the reliability of a transmission link, since it
provides a lowest bound on the transmission error probability \cite{1197843}. } for the considered backscatter system.
Assume that the $k$-th BD is selected to backscatter information to its receiver, while other backscatter links keep silent. Let $\mathcal{P}^b_{{\rm{out}},k}$ denote the outage probability for the $k$-th backscatter link. For a given SINR threshold $\gamma^b_{\rm{th}}$, $\mathcal{P}^b_{{\rm{out}},k}$ can be calculated as
\begin{align}\label{5}
\mathcal{P}^b_{{\rm{out}},k}=\mathbb{P}(E_k< P_{c,k} T)+\mathbb{P}(\gamma^b_k< \gamma^b_{\rm{th}}, E_k\geq P_{c,k} T),
\end{align}
where the first term denotes the energy outage probability that the $k$-th BD does not have enough energy for backscattering while the second term expresses the probability that the $k$-th BR fails to decode backscattered information in the case of $E_k\geq P_{c,k} T$.

\subsection{Adaptive RC Scheme}
For the backscatter link, we adopt an adaptive RC scheme to minimize the outage probability of the backscatter link.
Since minimizing the outage probability $\mathcal{P}_{{\rm{out}},k}^b$ is equivalent to maximizing the successful transmission probability for the backscatter link and the  successful transmission happens only when $\gamma^b_k\geq \gamma^b_{\rm{th}}$ and $E_k\geq P_{c,k} T$ hold simultaneously,
we can formulate the optimization problem as
\begin{align}\nonumber
\begin{array}{l}
{{\bf{P}}_1}:\;\mathop {{\rm{max}}}\limits_{{\beta _k}} \mathbb{P}(\gamma _k^b \ge \gamma _{{\rm{th}}}^b)\\
{\rm{s}}.{\rm{t}}.\;{\rm{C1}}:E_k\geq P_{c,k} T,\\
\;\;\;\;\;\;{\rm{C2}}:0 \le {\beta _k} \le 1,
\end{array}
\end{align}
where constraint $\rm{C1}$ should be satisfied to ensure that the harvested energy is enough for circuit operation consumption during backscattering. It is worth noting that for the case with $E_k< P_{c,k} T$, $\mathcal{P}_{{\rm{out}},k}^b$ is always equal to $1$ since the $k$-th BD is always inactive due to the lack of energy.

After some mathematical calculations, the problem ${{\bf{P}}_1}$ can be transformed as
\begin{align}\nonumber
\begin{array}{l}
{{\bf{P}}_2}:\;\mathop {{\rm{max}}}\limits_{{\beta _k}} \mathbb{P}(\gamma _k^b \ge \gamma _{{\rm{th}}}^b)\\
{\rm{s}}.{\rm{t}}.\;:\;\min\left( {\frac{{{\Phi _k}}}{{{P_t}{K_{1k}}|{h_{1k}}{|^2}}},1} \right) \le {\beta _k} \le 1,
\end{array}
\end{align}
where ${\Phi _k} = \frac{{\ln \frac{{{E_{\max ,k}}{e^{{s_{1k}}{s_{0k}}}} + {P_{c,k}}{e^{{s_{1k}}{s_{2k}}}}}}{{{E_{\max ,k}} - {P_{c,k}}}}}}{{{s_{1k}}}}$.

{\color{black} \emph{Proof.} Substituting \eqref{4aa} into  constraint $\rm{C}1$ and after  some mathematical manipulation, we have  ${P_{r,k}} \ge {\Phi _k}$. As ${P_{r,k}}=\beta_k{{P_t}{K_{1k}}{{\left| {{h_{1k}}} \right|}^2}}$ denotes the received power at the $k$-th BD, we can derive the following inequality, i.e.,
${\beta _k} \ge \frac{{{\Phi _k}}}{{{P_t}{K_{1k}}{{\left| {{h_{1k}}} \right|}^2}}}$.
Since $E_{{\rm{max}},k}>P_{c,k}$, ${\beta _k}>0$ always holds. Combining ${\beta _k} \ge \frac{{{\Phi _k}}}{{{s_{1k}}{P_t}{K_{1k}}{{\left| {{h_{1k}}} \right|}^2}}}$ with constraint $\rm{C}2$, we obtain the range of ${\beta _k}$ as $\min\left( {\frac{{{\Phi _k}}}{{{P_t}{K_{1k}}|{h_{1k}}{|^2}}},1} \right) \le {\beta _k} \le 1$.\hfill {$\blacksquare $}}

Since $\mathbb{P}(\gamma _k^b \ge \gamma _{{\rm{th}}}^b)$ decreases with the increasing of $\beta_k$, the optimal RC for the $k$-th BD is given by
\begin{align}\label{b1}
\beta_k^*=\min\left( {\frac{{{\Phi _k}}}{{{P_t}{K_{1k}}|{h_{1k}}{|^2}}},1} \right).
\end{align}
{\color{black}The derived expression in \eqref{b1} provides a guideline to set the value of RC in a practical AmBackCom.}

\subsection{Outage Analysis for the $k$-th Backscatter Link}
By applying $\beta_k^*$, $\mathcal{P}^b_{{\rm{out}},k}$ can be rewritten as
\begin{align}\label{8}\notag
\mathcal{P}^b_{{\rm{out}},k}&=\underbrace{\mathbb{P}\left(|{h_{1k}}{|^2} < \frac{{{b_k}}}{{{a_k}}}\right)}_{I_1}\\
&+\underbrace{\mathbb{P}\left(\frac{S_k}{{{P_t}{K_{k}^p}{{\left| {g_k^p} \right|}^2} + {\sigma ^2}}}<\gamma^b_{\rm{th}}, |{h_{1k}}{|^2} \geq \frac{{{b_k}}}{{{a_k}}}\right)\!}_{I_2},
\end{align}
where $a_k={\eta _k}{P_t}{K_{1k}}{K_{2k}}$, $b_k={\eta _k}{\Phi _k}{K_{2k}}$, and
${S_k} = \left\{ {\begin{array}{*{10}{c}}
{\!\!\!{a_k}|{h_{1k}}{|^2}|{h_{2k}}{|^2} - {b_k}|{h_{2k}}{|^2},\;{\rm{if}}\;{{\left| {{h_{1k}}} \right|}^2} \ge \frac{{{b_k}}}{{{a_k}}}}\\
{\!\!\!\!\!\!\!\!0,\;\;\;\;\;\;\;\;\;\;\;\;\;\;\;\;\;\;\;\;\;\;\;\;\;\;\;\;\;\;\;\;\;\;\;\;\;\;{\rm{otherwise}}}
\end{array}} \right.$.  

Since the multiplicative and additive  channel gain, ${S_k}$, is included in \eqref{8}, we first provide a proposition to derive its probability density function (PDF) and then calculate the outage probability of $k$-th backscatter link in the following context.

\begin{figure*}[t]
\normalsize
\setcounter{equation}{15}
\begin{align}\notag\label{12-1}
\mathcal{P}^{\rm{HS}}_{{\rm{out}},k}&\approx 1\! -\! \exp \left( { - \frac{{{b_k}}}{{{a_k}{\lambda _{1k}}}}} \right)\bigg[1 - {F_{{S_k}}}\left( {\gamma _{{\rm{th}}}^b{\sigma ^2}} \right) \!-\! \exp \left( {\frac{{{\sigma ^2}}}{{{P_t}K_k^p\lambda _k^p}}} \right) \bigg[ {\Theta _k} - \frac{1}{{{\lambda _{1k}}{\lambda _{2k}}{a_k}}}\bigg( \frac{{{{ {\gamma _{{\rm{th}}}^b} }}{\sigma ^4}}}{{{P_t}K_k^p\lambda _k^p}}\left( {\ln \left( {\frac{{\sqrt {\gamma _{{\rm{th}}}^b} \sigma }}{{\sqrt {{\lambda _{1k}}{\lambda _{2k}}{a_k}} }}} \right) \!-\! \frac{1}{4}} \right) \\
&- 2\gamma _{{\rm{th}}}^b{\sigma ^2}\left( {\ln \left( {\frac{{\sqrt {\gamma _{{\rm{th}}}^b} \sigma }}{{\sqrt {{\lambda _{1k}}{\lambda _{2k}}{a_k}} }}} \right) - \frac{1}{2}} \right) \bigg) + 2\gamma _{{\rm{th}}}^b{\vartheta _k}{c_0}\left( {1 - \exp \left( { - \frac{{{\sigma ^2}}}{{{P_t}K_k^p\lambda _k^p}}} \right)} \right) \bigg]\bigg].
\end{align}
\setcounter{equation}{11}
\hrulefill
\end{figure*}

\emph{\textbf{Proposition 1:}} Based on $|{h_{2k}}{|^2}\sim \exp(\frac{1}{\lambda_{2k}})$, conditioning on $|{h_{1k}}{|^2} \!\!\geq\! \frac{{{b_k}}}{{{a_k}}}$, the cumulative distribution function (CDF) of $S_k$, denoted by $F_{S_k}(x)$, and the PDF of $S_k$, denoted by $f_{S_k}(x)$, are, respectively, given by
\begin{align}\label{10}
F_{S_k}(x)=1 - \frac{1}{{{\lambda _{1k}}{a_k}}}\sqrt {\frac{{4{a_k}{\lambda _{1k}}x}}{{{\lambda _{2k}}}}} {K_1}\left( {\sqrt {\frac{{4x}}{{{\lambda _{2k}}{\lambda _{1k}}{a_k}}}} } \right),
\end{align}
\begin{align}
f_{S_k}(x)=\frac{2}{{{\lambda _{1k}}{\lambda _{2k}}{a_k}}}{K_0}\left( {2\sqrt {\frac{x}{{{\lambda _{1k}}{\lambda _{2k}}{a_k}}}} } \right),
\end{align}
where both $K_0(\cdot)$ are $K_1(\cdot)$ are the modified Bessel functions of the second kind \cite{b1}.

\emph{Proof:} See Appendix A.\hfill {$\blacksquare $}

Using Proposition 1, we can obtain the following theorem.

\emph{Theorem 1.} The outage probability of the $k$-th backscatter link can be calculated as  
\begin{align}\notag \label{a10}
&\mathcal{P}^b_{{\rm{out}},k}=\\ \notag
&1 - \exp \left( { - \frac{{{b_k}}}{{{a_k}{\lambda _{1k}}}}} \right)\bigg[ 1 - {F_{{S_k}}}\left( {\gamma _{{\rm{th}}}^b{\sigma ^2}} \right)-\exp \left( {\frac{{{\sigma ^2}}}{{{P_t}K_k^p\lambda _k^p}}} \right) \\
& \times \left( {{\Theta _k} \!- \! \int_0^{\gamma _{{\rm{th}}}^b{\sigma ^2}} {\exp } \left( { - \frac{x}{{\gamma _{{\rm{th}}}^b{P_t}K_k^p\lambda _k^p}}} \right){f_{{S_k}}}\left( x \right)dx} \right)\! \bigg],
\end{align}
where  $\Theta_k=- {\vartheta _k}\gamma _{{\rm{th}}}^b{\rm{Ei}}( - {\vartheta _k}\gamma _{{\rm{th}}}^b)\exp \left( {{\vartheta _k}\gamma _{{\rm{th}}}^b} \right)$ with ${\vartheta _k}=\frac{{{P_t}{K_{k}^p}{\lambda _{k}^p}}}{{{\lambda _{1k}}{\lambda _{2k}}{a_k}}}$ and the exponential integral function ${\rm{Ei}}(\varrho)=\int_{-\infty}^{\varrho}t^{-1}e^{t}dt$ [eq. (8.211.1), \cite{b1}].

\emph{Proof:} See Appendix B.\hfill {$\blacksquare $}

Although \eqref{a10} is more analytical than  \eqref{8}, there is still no closed-form expression for $\mathcal{P}^b_{{\rm{out}},k}$ due to the involved integral $\int_{0}^{ \gamma _{{\rm{th}}}^b{\sigma ^2}}  \exp \left( { - \frac{x}{{\gamma _{{\rm{th}}}^b{P_t}{K_{k}^p}{\lambda _{k}^p}}}} \right){f_{{S_k}}}\left( {x} \right)dx$.  In what follows, we provide two ways to address this problem. One is to use Gaussian-Chebyshev quadrature to obtain an approximation for $\mathcal{P}^b_{{\rm{out}},k}$ for any given $P_t$, since it  can provide sufficient level of accuracy within very few terms \cite{8633928,8613860,7445146}. The other is to approximate $\mathcal{P}^b_{{\rm{out}},k}$ based on a high transmit power of the LT. Such an approach aims to find some insights  and has been widely used in the field of performance analysis.

 \emph{Gaussian-Chebyshev Approximation:} Based on the Gaussian-Chebyshev quadrature, we have the following approximation, i.e.,
 \begin{align}  \notag
 {\cal P}_{{\rm{out}},k}^b &\approx 1 - \exp \left( { - \frac{{{b_k}}}{{{a_k}{\lambda _{1k}}}}} \right)\\ \notag
 & \times \bigg[ 1 - {F_{{S_k}}}\left( {\gamma _{{\rm{th}}}^b{\sigma ^2}} \right) - \exp \left( {\frac{{{\sigma ^2}}}{{{P_t}K_k^p\lambda _k^p}}} \right) \\  \notag
     & \times \bigg( {\Theta _k} - \frac{{\pi \gamma _{{\rm{th}}}^b{\sigma ^2}}}{{2M}} \sum\limits_{m = 1}^M  \sqrt {1 - v_m^2}
 \end{align}
 \begin{align} \label{12}
 &\times\exp \left( { - \frac{{\kappa _m^{(0)}}}{{\gamma _{{\rm{th}}}^b{P_t}K_k^p\lambda _k^p}}} \right) {f_{{S_k}}}\left( {\kappa _m^{(0)}} \right) \bigg) \bigg],
 \end{align}
where ${v_m} = \cos \frac{{2m - 1}}{{2M}}\pi $, ${\kappa _m^{(0)}} = \frac{\gamma _{{\rm{th}}}^b{\sigma ^2 }}{2}{v_m} + \frac{\gamma _{{\rm{th}}}^b{\sigma ^2 }}{2}$, and $M$ is a parameter that determines the tradeoff between complexity and accuracy.

{\color{black}\emph{Remark 1:} The derived expression  in \eqref{12} has  the following applications. Firstly, it provides a closed-form expression that can accurately evaluate  the outage probability of the $k$-th backscatter link  with a small $M$,  thus avoiding the necessity  of Monte Carlo simulations. This would be useful   for the service providers and the industry, who may take up experiments and/or implementations based on their assessments of the reported results, because they need to know the bound performance of a practical system. Secondly, the outage probability expression in \eqref{12} and the numerical results generated using it enable us to obtain useful insights into how the system parameters affect the outage performance of a backscatter system.}

 \emph{Outage Probability with a High Transmit Power of the LT: }
Assuming a high transmit power of the LT, we derive  the outage probability for the $k$-th backscatter link, denoted by $\mathcal{P}^{\rm{HS}}_{{\rm{out}},k}$, which is given  \eqref{12-1}, as shown at the top of the next page.

\emph{Proof:} See Appendix D.\hfill {$\blacksquare $}

\setcounter{equation}{16}
Using \eqref{12-1}, we can  see that  $\mathcal{P}^{\rm{HS}}_{{\rm{out}},k}$  converges to a lower bound when $P_t \to \infty$, given by
\begin{align}\notag\label{17a}
\mathop {\lim }\limits_{{P_t} \to \infty } {\cal P}_{{\rm{out}},k}^{{\rm{HS}}}\!& =  -\frac{{K_k^p\lambda _k^p\gamma _{{\rm{th}}}^b}}{{{\lambda _{1k}}{\lambda _{2k}}{\eta _k}{K_{1k}}{K_{2k}}}} {\rm{Ei}}\! \left( \!{ - \frac{{K_k^p\lambda _k^p\gamma _{{\rm{th}}}^b}}{{{\lambda _{1k}}{\lambda _{2k}}{\eta _k}{K_{1k}}{K_{2k}}}}} \right)\\
&\times \exp \left( {\frac{{K_k^p\lambda _k^p\gamma _{{\rm{th}}}^b}}{{{\lambda _{1k}}{\lambda _{2k}}{\eta _k}{K_{1k}}{K_{2k}}}}} \right).
\end{align}

\emph{Proof:} If $P_t \to \infty$, ${\vartheta _k}$  is bound and $a_k$ converges to infinity. Thus, $\exp \left( { - \frac{{{b_k}}}{{{a_k}{\lambda _{1k}}}}} \right)$, ${F_{{S_k}}}\left( {\gamma _{{\rm{th}}}^b{\sigma ^2}} \right)$, $\exp \left( {\frac{{{\sigma ^2}}}{{{P_t}K_k^p\lambda _k^p}}} \right)$ and ${\exp \left( { - \frac{{{\sigma ^2}}}{{{P_t}K_k^p\lambda _k^p}}} \right)}$ converge to one. Combing this conclusion, i.e., $\mathop {\lim }\limits_{{a_k} \to \infty } \frac{{\ln \left( {\frac{{\sqrt {\gamma _{{\rm{th}}}^b} \sigma }}{{\sqrt {{\lambda _{1k}}{\lambda _{2k}}{a_k}} }}} \right)}}{{{\lambda _{1k}}{\lambda _{2k}}{a_k}}} = \mathop {\lim }\limits_{{a_k} \to \infty } \frac{{ - \frac{1}{{2{a_k}}}}}{{{\lambda _{1k}}{\lambda _{2k}}}} = 0$, \eqref{17a} can be obtained. \hfill {$\blacksquare $}

\emph{Remark 2:} The bound in \eqref{17a} reveals that the interference caused by the legacy transmission leads to the  outage saturation phenomenon of the $k$-th backscatter link, and that the diversity gain of the $k$-th backscatter link can be calculated as $\mathop {\lim }\limits_{{P_t} \to \infty }  - \frac{{\log \left( {{\cal{P}}_{{\rm{out}},k}^{{\rm{HS}}}} \right)}}{{\log {P_t}}} = 0$. This is because the interference link from the LT-to-the $k$-th BR scales along the transmit power of the LT at each block, which leads to a bound   of ${\gamma^b _k}$ as $P_t$ grows.
Also, it  reveals that how the  channel qualities impact  on
the low bound for the outage probability of the $k$-th backscatter link. In particular, $\mathop {\lim }\limits_{{P_t} \to \infty } {\cal P}_{{\rm{out}},k}^{{\rm{HS}}}$ decreases with the increase of $\frac{{K_k^p\lambda _k^p}}{{{\lambda _{1k}}{\lambda _{2k}}{K_{1k}}{K_{2k}}}}$ that is related with the means of the channel gains of the LT-to-the $k$-th BD link, the $k$-th backscatter link and the interfering link from LT-to-BR.

To better understand the performance bounds of backscatter systems,  we turn our attention to investigate its outage probability  under two extreme cases.
\subsubsection{Outage Probability under the Best Case}
Specifically, the selected backscatter link with the maximum SINR at the BR results in the best case.
Denote the outage probability of this case as $\mathcal{P}_{{\rm{out}},b}^{\rm{bc}}$, which can be computed as
\begin{align}\label{13}\notag
\mathcal{P}_{{\rm{out}},b}^{\rm{bc}}&\!=\!\!\mathbb{P}\!\!\left(\!\!\max \!\!\left( \! {\frac{{{S_1}}}{{{P_t}{K_{r1}}{{\left| {g_1^p} \right|}^2}\!\! +\!\! {\sigma ^2}}}, \ldots ,\frac{{{S_K}}}{{{P_t}{K_{rK}}{{\left| {g_K^p} \right|}^2} \!\!+\!\! {\sigma ^2}}}}\! \right)\!\! <\!\! \gamma _{{\rm{th}}}^b\!\!\right)\\ \notag
&\overset{\text{(a)}}{=}\prod\limits_{k = 1}^K \mathbb{P}{\left( {\frac{{{S_k}}}{{{P_t}{K_{k}^p}{{\left| {g_k^p} \right|}^2} + {\sigma ^2}}} < \gamma _{{\rm{th}}}^b} \right)} \!\!\!\!\\
&=\prod\limits_{k = 1}^K {{\cal{P}}_{{\rm{out}},k}^b},
\end{align}
where step (a) holds since {\small{${\left\{ {\frac{{{S_k}}}{{{P_t}{K_{k}^p}{{\left| {g_k^p} \right|}^2} + {\sigma ^2}}}} \right\}_{k \in \left\{ {1, \ldots ,K} \right\}}}$}} are independent of each other.

Accordingly, the lower bound of the outage probability for this case is given as
\begin{align}\label{19a}
\mathop {{\rm{lim}}}\limits_{{P_t} \to \infty } {\cal P}_{{\rm{out}},b}^{{\rm{bc}}} = \prod\limits_{k = 1}^K\mathop {\lim }\limits_{{P_t} \to \infty } {\cal{P}}_{{\rm{out}},k}^{{\rm{HS}}}.
\end{align}

\subsubsection{Outage Probability under the Worst Case}
The worst case will happen when the selected backscatter link achieves the minimum SINR at the BR.
Under such case, the outage probability, denoted by $\mathcal{P}_{{\rm{out}},b}^{\rm{wc}}$, is given by
\begin{align}\notag \label{14}
&\mathcal{P}_{{\rm{out}},b}^{\rm{wc}}\\ \notag
&=\mathbb{P}\!\left(\!\!\min \!\left( \! {\frac{{{S_1}}}{{{P_t}{K_{r1}}{{\left| {g_1^p} \right|}^2}\! +\! {\sigma ^2}}}, \ldots ,\frac{{{S_K}}}{{{P_t}{K_{rK}}{{\left| {g_K^p} \right|}^2} \!\!+\!\! {\sigma ^2}}}}\! \right)\!\! <\!\! \gamma _{{\rm{th}}}^b\!\!\right) \\ \notag
&=1\!-\!\mathbb{P}\!\!\left(\!\!\min \!\!\left( \! {\frac{{{S_1}}}{{{P_t}{K_{r1}}{{\left| {g_1^p} \right|}^2}\!\! +\!\! {\sigma ^2}}}, \ldots ,\frac{{{S_K}}}{{{P_t}{K_{rK}}{{\left| {g_K^p} \right|}^2} \!\!+\!\! {\sigma ^2}}}}\! \right)\!\! \geq\!\! \gamma _{{\rm{th}}}^b\!\!\right)\\ \notag
&=1-\prod\limits_{k = 1}^K {\left( {\frac{{{S_k}}}{{{P_t}{K_{k}^p}{{\left| {g_k^p} \right|}^2} + {\sigma ^2}}} \ge \gamma _{{\rm{th}}}^b} \right)}\\
&=1-\prod\limits_{k = 1}^K \left(1-{{\cal{P}}_{{\rm{out}},k}^b}\right).
\end{align}

Similarly as \eqref{19a}, the low bound of the outage probability in the worst case for the backscatter system is written as
\begin{align}
\mathop {{\rm{lim}}}\limits_{{P_t} \to \infty } {\cal P}_{{\rm{out}},b}^{{\rm{wc}}} = 1 - \prod\limits_{k = 1}^K {\left( {1 - \mathop {\lim }\limits_{{P_t} \to \infty } {\cal{P}}_{{\rm{out}},k}^{{\rm{HS}}}} \right)}.
\end{align}
\section{Outage Analysis for Legacy Transmission}
In this section, we study the outage performance of the legacy system in order to see how backscatter system influences the legacy transmission. 
For the legacy transmission with a given backscatter link, an outage event will occur in two cases. The first case is that the  BD does not have enough energy for backscattering while the received SNR at the LR is less than the given threshold. The second case is that the  BD has enough energy for backscattering while the received SINR at the LR is less than the given threshold.
Let $\mathcal{P}^l_{{\rm{out}},k}$ denote the outage probability for the legacy link when the $k$-th backscatter link is selected. Let $\gamma_{\rm{th}}$ be the threshold for the legacy transmission. Then $\mathcal{P}^l_{{\rm{out}},k}$ is given by
\begin{align}\notag \label{6}
\mathcal{P}^l_{{\rm{out}},k}&=\mathbb{P}(\gamma^{(1)}_k<\gamma_{\rm{th}},E_k\geq P_{c,k} T)\\
&+\mathbb{P}(\gamma^{(2)}_k<\gamma_{\rm{th}},\!E_k< P_{c,k} T),
\end{align}
where the first term and the second term denote the outage probabilities for the cases with $E_k\!\!\geq\! P_{c,k} T$ and $E_k\!\!<\! P_{c,k} T$, respectively.

Considering $\beta_k^*$, $\mathcal{P}^l_{{\rm{out}},k}$ can be rewritten as
\begin{align}\label{15}\notag
\mathcal{P}^l_{{\rm{out}},k}=& \underbrace{\mathbb{P}\left(\frac{{{P_t}{K_p}{{\left| {{h_p}} \right|}^2}}}{{\sigma ^2}} \!<\! {\gamma _{{\rm{th}}}},|{h_{1k}}{|^2}\! <\! \frac{{{b_k}}}{{{a_k}}}\right)}_{I_3}\\
&+\underbrace{\mathbb{P}\left(\frac{{{P_t}{K_p}|{h_p}{|^2}}}{{{G_k} + {\sigma ^2}}}\! <\! {\gamma _{{\rm{th}}}},|{h_{1k}}{|^2} \!\ge\! \frac{{{b_k}}}{{{a_k}}}\right)}_{I_4} ,
\end{align}
where 
$a_{ik}={\eta _k}{P_t}{K_{1k}}{K_{k}^s}$, $b_{ik}={\eta _k}{\Phi _k}{K_{k}^s}$, and ${G_k} = \left\{ {\begin{array}{*{20}{c}}
{\!\!\!{a_{ik}}|{h_{1k}}{|^2}|g_k^s{|^2} - {b_{ik}}|g_k^s{|^2},\;{\rm{if}}\;|{h_{1k}}{|^2} \ge \frac{{{b_k}}}{{{a_k}}}}\\
{\!\!0,\;\;\;\;\;\;\;\;\;\;\;\;\;\;\;\;\;\;\;\;\;\;\;\;\;\;\;\;\;\;\;\;\;\;\;\;\;\;\;\;\;{\rm{otherwise}}}
\end{array}} \right.$.

Since both ${\left| {{h_p}} \right|}^2$ and $|{h_{1k}}{|^2}$ are independent of each other, $I_3$ can be calculated as
\begin{align}\label{16}\notag
I_3&=\mathbb{P}\left( {\frac{{{P_t}{K_p}{{\left| {{h_p}} \right|}^2}}}{{{\sigma ^2}}} < {\gamma _{{\rm{th}}}}} \right)\mathbb{P}\left( {|{h_{1k}}{|^2} < \frac{{{b_k}}}{{{a_k}}}} \right)\\
&=\!\left(\! {1 \!-\! \exp \left( { - \frac{{{\gamma _{{\rm{th}}}}{\sigma ^2}}}{{{P_t}{K_p}{\lambda _p}}}} \right)}\! \right)\!\left( {1 - \exp \left( { - \frac{{{b_k}}}{{{\lambda _{1k}}{a_k}}}} \right)} \!\right).\!\!
\end{align}

Similarly to the derivation in Appendix B,  $I_4$ can be written as
\begin{align}\label{17}\notag
I_4&=\mathbb{P}\left( {|{h_p}{|^2} < \frac{{{\gamma _{{\rm{th}}}}\left( {{G_k} \!+\! {\sigma ^2}} \right)}}{{{P_t}{K_p}}}\bigg||{h_{1k}}{|^2} \!\ge \!\frac{{{b_k}}}{{{a_k}}}} \!\right)\!\mathbb{P}\!\left(\! {|{h_{1k}}{|^2} \!\ge \!\frac{{{b_k}}}{{{a_k}}}} \right)\\
&=\exp\left(-\frac{b_k}{a_k\lambda_{1k}}\right)\left[1 - \exp \left( { - \frac{{{\gamma _{{\rm{th}}}}{\sigma ^2}}}{{{P_t}{K_p}{\lambda _p}}}} \right)I_5\right],
\end{align}
where $I_5=\int_0^{ + \infty } {\exp \left( { - \frac{{{\gamma _{{\rm{th}}}}x}}{{{P_t}{K_p}{\lambda _p}}}} \right)} {f_{{G_k}}}\left( x \right)dx$ and ${f_{{G_k}}}\left( x \right)$ denotes the PDF of $G_k$ under the case of $|{h_{1k}}{|^2} \!\!\geq\! \frac{{{b_k}}}{{{a_k}}}$.

Similarly as Proposition 1, we have ${f_{{G_k}}}\left( x \right) = \frac{2}{{{\lambda _{1k}}\lambda _k^s{a_{ik}}}}{K_0}\left( {2\sqrt {\frac{x}{{{\lambda _{1k}}\lambda _k^s{a_{ik}}}}} } \right)$.
Based on Appendix C, $I_5$ can be computed as
\begin{align}\label{18}
I_5= - {\Xi _k}\exp \left( {{\Xi _k}} \right){\rm{Ei}}\left( { - {\Xi _k}} \right),
\end{align}
where ${\Xi _k} = \frac{{{P_t}{K_p}{\lambda _p}}}{{{\gamma _{\rm{th}}}{\lambda _{1k}}\lambda _k^s{a_{ik}}}}$.

Substituting \eqref{16}, \eqref{17} and \eqref{18} into \eqref{15}, $\mathcal{P}^l_{{\rm{out}},k}$ is obtained. Then we can obtain the lower bound for the outage probability of the legacy transmission, given by
\begin{align}\notag
& \mathop {\lim }\limits_{{P_t} \to \infty } {\cal P}_{{\rm{out}},k}^l = 1 + \frac{{{K_p}{\lambda _p}}}{{{\eta _k}{\gamma _{{\rm{th}}}}{\lambda _{1k}}\lambda _k^s{K_{1k}}K_k^s}}\times \\
& \exp \left( {\frac{{{K_p}{\lambda _p}}}{{{\eta _k}{\gamma _{{\rm{th}}}}{\lambda _{1k}}\lambda _k^s{K_{1k}}K_k^s}}} \right){\rm{Ei}}\left( { - \frac{{{K_p}{\lambda _p}}}{{{\eta _k}{\gamma _{{\rm{th}}}}{\lambda _{1k}}\lambda _k^s{K_{1k}}K_k^s}}} \right).
\end{align}

In the following, we will further analyze the outage probability for the legacy transmission under two extreme cases so that we can know the bound performance of the legacy transmission. The first case (also termed as best case) is caused by
the selected backscatter link with the minimum interference on the LR and will lead to the lowest outage probability of the legacy transmission.
The second case is the worst case, where the selected backscatter link results in the maximum interference on the LR.

\subsection{Outage Probability under the Best Case}
There are two cases for the outage probability under the best case. Specifically, when there is at least one backscatter link that is inactive due to the lack of energy, the best case for the legacy transmission happens with the inactive backscatter link selected. In this case, the received SNR at the LR is given by ${\gamma ^{(2)}_k}$. When all the backscatter links have enough energy for backscattering, the backscatter link with the minimum interference ${G_{\min }}=\min\{G_1,G_2,\cdots, G_K\}$ should be selected and the received SINR is determined by ${\gamma ^{l}_{\rm{bc}}}=\!\frac{{{P_t}{K_p}{{\left| {{h_p}} \right|}^2}}}{{{G_{\min }}+\sigma ^2}}$.
Accordingly, we can compute the outage probability under the best case as
\begin{align}\label{20}
\mathcal{P}_{{\rm{out}},l}^{\rm{bc}}=\mathcal{P}_{{\rm{out}},l}^{\rm{No-SL}}(1-\mathcal{P}_1)+\mathcal{P}_2\mathcal{P}_1,
\end{align}
where $\mathcal{P}_1$ is the probability that all the backscatter links have enough energy for backscattering; $\mathcal{P}_2$ is the probability of ${\gamma ^{l}_{\rm{bc}}}<\! {\gamma _{{\rm{th}}}}$ conditioning on all active backscatter links, and  $\mathcal{P}_{{\rm{out}},l}^{\rm{No-SL}}$ is the outage probability of the legacy transmission when the backscatter link is inactive, i.e., the backscatter link has no impacts on the legacy transmission.  $\mathcal{P}_{{\rm{out}},l}^{\rm{No-SL}}$ is given by
\begin{align}\label{19} \notag
\mathcal{P}_{{\rm{out}},l}^{\rm{No-SL}}&=\mathbb{P}\left(\frac{{{P_t}{K_p}{{\left| {{h_p}} \right|}^2}}}{{\sigma ^2}} < {\gamma _{{\rm{th}}}}\!\right)\\
&=1-\exp \left( { - \frac{{{\gamma _{{\rm{th}}}}{\sigma ^2}}}{{{P_t}{K_p}{\lambda _p}}}} \right).
\end{align}

Thus, in order to obtain the value of $\mathcal{P}_{{\rm{out}},l}^{\rm{bc}}$, we should determine the values of $\mathcal{P}_1$ and $\mathcal{P}_2$ first.
Based on the definitions, $\mathcal{P}_1$ can be calculated as
\begin{align}\label{20_1}
\mathcal{P}_1=\mathbb{P}\left(\bigcap\limits_{k = 1}^K {|{h_{1k}}{|^2} > \frac{{{b_k}}}{{{a_k}}}}\right)=\prod\limits_{k = 1}^K {\exp \left( { - \frac{{{b_k}}}{{{\lambda _{1k}}{a_k}}}} \right)}.
\end{align}
$\mathcal{P}_2$ is determined by
\begin{align}\label{20_2}\notag
\mathcal{P}_2&=\mathbb{P}\left({\gamma ^{l}_{\rm{bc}}}< {\gamma _{{\rm{th}}}}\bigg|\bigcap\limits_{k = 1}^K {|{h_{1k}}{|^2} > \frac{{{b_k}}}{{{a_k}}}}\right)\\ \notag
&=1 - \exp \left( { - \frac{{{\gamma _{{\rm{th}}}}{\sigma ^2}}}{{{P_t}{K_p}{\lambda _p}}}} \right) \\
&\times {\mathbb{E}}\left[ {\exp \left( { - \frac{{{\gamma _{{\rm{th}}}}{G_{\min }}}}{{{P_t}{K_p}{\lambda _p}}}}\right)}\bigg|\bigcap\limits_{k = 1}^K {|{h_{1k}}{|^2} >  \frac{{{b_k}}}{{{a_k}}}}\right].
\end{align}

\emph{\textbf{Proposition 2:}} The CDF of $\min \left( {{G_1},{G_2}, \ldots ,{G_K}} \right)$ conditioned on $\bigcap\limits_{k = 1}^K {|{h_{1k}}{|^2} > \frac{{{b_k}}}{{{a_k}}}}$, denoted by $F_{G_{\min}}(x)$, is given by
\begin{align}\label{21}\notag
&F_{G_{\min}}(x)\\
&=1-\prod\limits_{k = 1}^K  \left[ \!{\frac{{1}}{{{\lambda _{1k}}{a_{ik}}}}\!\sqrt {\frac{{4{a_{ik}}{\lambda _{1k}}x}}{{\lambda _k^s}}} {K_1}\left({\sqrt {\frac{{4x}}{{{\lambda _{1k}}\lambda _k^s{a_{ik}}}}} } \right)} \right].
\end{align}

\emph{Proof:} See Appendix E.\hfill {$\blacksquare $}

According to the Proposition 2, $\mathcal{P}_{2}$ can be computed as
\begin{align}\label{22}\notag
&\mathcal{P}_{2}\\ \notag
&=1 - \exp \left(\! { - \frac{{{\gamma _{{\rm{th}}}}{\sigma ^2}}}{{{P_t}{K_p}{\lambda _p}}}} \!\right)\!\int_0^{ + \infty } {\exp\! \left(\! { - \frac{{{\gamma _{{\rm{th}}}}x}}{{{P_t}{K_p}{\lambda _p}}}} \!\right)}\! {f_{{G_{\min }}}}\!\left( x \right)\!dx \\ \notag
&\overset{\text{(a)}}{=}1 - {e^{ - \frac{{{\gamma _{{\rm{th}}}}{\sigma ^2}}}{{{P_t}{K_p}{\lambda _p}}}}}\int_0^{ + \infty }  \exp \left( { - t} \right){F_{{G_{\min }}}}\left( {\frac{{{P_t}{K_p}{\lambda _p}t}}{{{\gamma _{{\rm{th}}}}}}} \right)dt\!\\
&\overset{\text{(b)}}{=}1 - {e^{ - \frac{{{\gamma _{{\rm{th}}}}{\sigma ^2}}}{{{P_t}{K_p}{\lambda _p}}}}}\int_0^1 {{F_{{G_{\min }}}}\left( { - \frac{{{P_t}{K_p}{\lambda _p}}}{{{\gamma _{{\rm{th}}}}}}\ln y} \right)} dy,
\end{align}
where ${f_{{G_{\min }}}}\left( x \right)$ is the PDF of $G_{\min}$ conditioned on $\bigcap\limits_{k = 1}^K {|{h_{1k}}{|^2} > \frac{{{b_k}}}{{{a_k}}}}$; step (a) follows by using the subsection integral method and letting $t = \frac{{{\gamma _{{\rm{th}}}}x}}{{{P_t}{K_p}{\lambda _p}}}$;
step (b) holds by letting $y=e^{-t}$.

Similar to \eqref{12}, we employ Gaussian-Chebyshev quadrature to obtain an approximation for $\mathcal{P}_{2}$, given by
\begin{align}\label{22.1}\notag
\mathcal{P}_{2}&\approx1 - \frac{{\pi {e^{ - \frac{{{\gamma _{{\rm{th}}}}{\sigma ^2}}}{{{P_t}{K_p}{\lambda _p}}}}}}}{{2M}}\\
&\times \sum\limits_{m = 1}^M {\sqrt {1 - v_m^2} } {F_{{G_{\min }}}}\left( { - \frac{{{P_t}{K_p}{\lambda _p}}}{{{\gamma _{{\rm{th}}}}}}\ln\kappa _m^{(1)}} \right),
\end{align}
where ${\kappa _m^{(1)}} = \frac{v_m}{2} + \frac{1 }{2}$.

Based on \eqref{20}, \eqref{19}, \eqref{20_1} and \eqref{22.1}, ${\cal P}_{{\rm{out}},l}^{{\rm{bc}}}$ can be obtained. Accordingly, the lower bound for the outage probability of the legacy transmission in the best case can be written as
\begin{align}\notag
\mathop {{\rm{lim}}}\limits_{{P_t} \to \infty } {\cal P}_{{\rm{out}},l}^{{\rm{bc}}} &\approx 1 - \frac{\pi }{{2M}}\sum\limits_{m = 1}^M {\sqrt {1 - v_m^2} }\\
& \times {F_{{G_{\min }}}}\left( { - \frac{{{P_t}{K_p}{\lambda _p}}}{{{\gamma _{{\rm{th}}}}}}\ln \kappa _m^{(1)}} \right).
\end{align}
\subsection{Outage Probability under the Worst Case}
Likewise, the outage probability under the worst case is given by
\begin{align}\label{23}\notag
&\mathcal{P}_{{\rm{out}},l}^{\rm{wc}}\\ \notag
&=1\!-\! \exp\! \left(\! { - \frac{{{\gamma _{{\rm{th}}}}{\sigma ^2}}}{{{P_t}{K_p}{\lambda _p}}}}\! \right)\!{\mathbb{E}}\!\left[\! {\exp \!\left( { - \frac{{{\gamma _{{\rm{th}}}}{\widehat{G}_{\max }}}}{{{P_t}{K_p}{\lambda _p}}}} \!\right)} \right]\\ \notag
&=1 \!- \!\exp \left( { - \frac{{{\gamma _{{\rm{th}}}}{\sigma ^2}}}{{{P_t}{K_p}{\lambda _p}}}} \right)\!\!\int_0^{ + \infty }\!\!\!\!\!\!  \exp \left( { - \frac{{{\gamma _{{\rm{th}}}}x}}{{{P_t}{K_p}{\lambda _p}}}} \right){f_{{{\widehat G}_{\max }}}}\left( x \right)dx\\
&=1 \!- \!\exp \!\left(\! { - \frac{{{\gamma _{{\rm{th}}}}{\sigma ^2}}}{{{P_t}{K_p}{\lambda _p}}}} \!\right)\int_0^1 \! {F_{{{\widehat G}_{\max }}}}\left(\!\! { - \frac{{{P_t}{K_p}{\lambda _p}}}{{{\gamma _{{\rm{th}}}}}}\ln t} \right)dt,
\end{align}
where ${\widehat{G}_{\max }}=\max\{\widehat{G}_1,\widehat{G}_2,\cdots, \widehat{G}_K\}$ with ${{\widehat G}_k} = \left\{ {\begin{array}{*{20}{c}}
{{G_k},|{h_{1k}}{|^2} > \frac{{{b_k}}}{{{a_k}}}}\\
{0,|{h_{1k}}{|^2} \le \frac{{{b_k}}}{{{a_k}}}}
\end{array}} \right.$, ${f_{{{\widehat G}_{\max }}}}\left( x \right)$ denotes the PDF of ${{\widehat G}_{\max }}$ and  ${F_{{{\widehat G}_{\max }}}}\left( x \right)$ is the CDF of ${{\widehat G}_{\max }}$.
\begin{figure}
  \centering
  \includegraphics[width=0.4\textwidth]{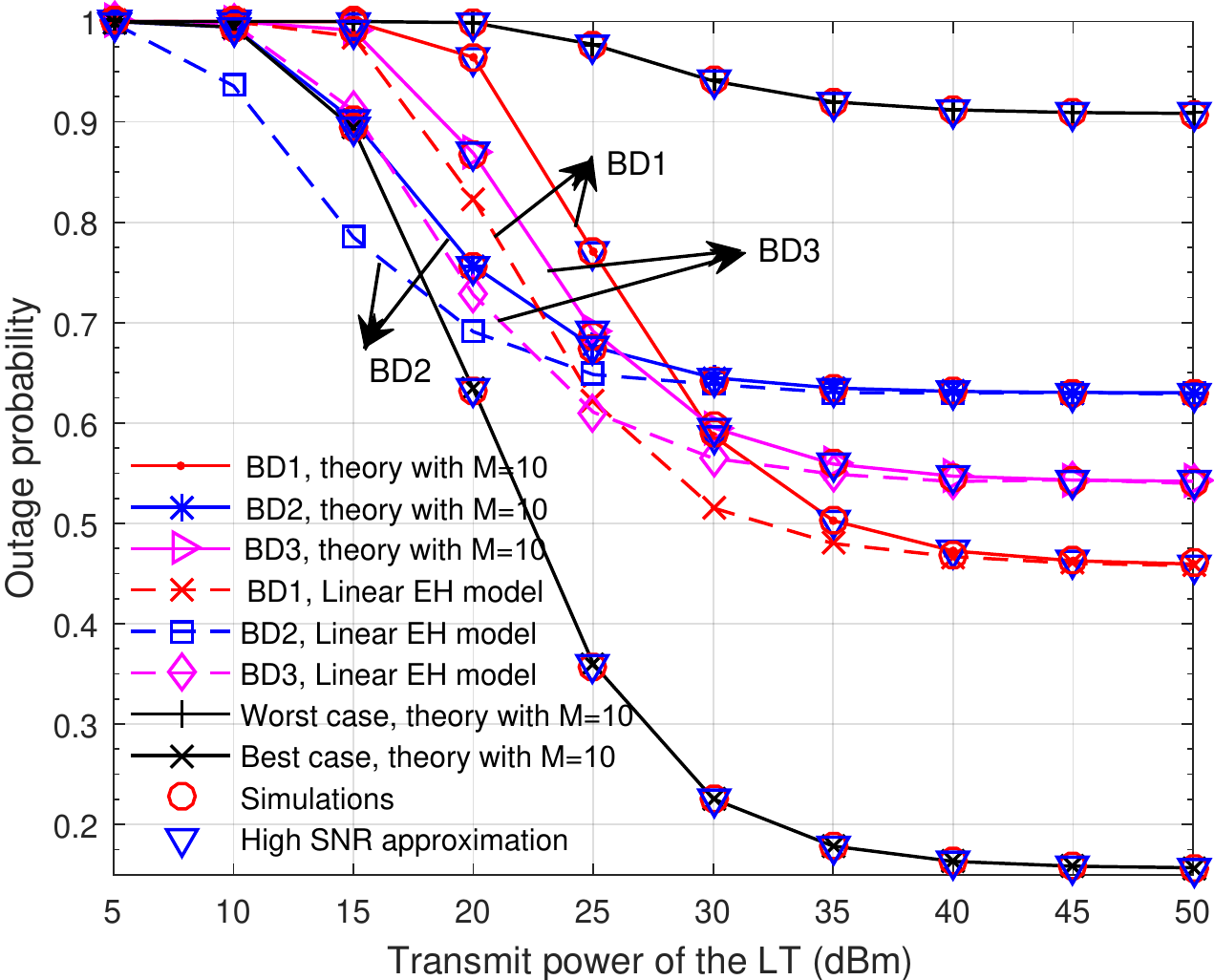}\\
  \caption{Outage probability for a backscatter system versus the transmit power of the LT.}\label{fig1}
\end{figure}

\emph{\textbf{Proposition 3:}} The CDF of $\max\{\widehat{G}_1,\widehat{G}_2,\cdots, \widehat{G}_K\}$, denoted by $F_{\widehat{G}_{\max}}(x)$, is given by
\begin{align}\label{24}\notag
&F_{\widehat{G}_{\max}}(x)\\ \notag
&=\prod\limits_{k = 1}^K \left( {1 - \frac{1}{{{\lambda _{1k}}{a_{ik}}}}\sqrt {\frac{{4{a_{ik}}{\lambda _{1k}}x}}{{\lambda _k^s}}} {K_1}\left( {\sqrt {\frac{{4x}}{{{\lambda _{1k}}\lambda _k^s{a_{ik}}}}} } \right)} \right)\\
&\times \exp \left( { - \frac{{{b_k}}}{{{a_k}{\lambda _{1k}}}}} \right) + 1 - \exp \left( { - \frac{{{b_k}}}{{{a_k}{\lambda _{1k}}}}} \right).
\end{align}

\emph{Proof:} See Appendix F.\hfill {$\blacksquare $}

Based on the Proposition 3, by using Gaussian-Chebyshev quadrature, $\mathcal{P}_{{\rm{out}},l}^{\rm{wc}}$ can be approximated as
\begin{align}\label{25}\notag
\mathcal{P}_{{\rm{out}},l}^{\rm{wc}}&\approx1 - \frac{{\pi {e^{ - \frac{{{\gamma _{{\rm{th}}}}{\sigma ^2}}}{{{P_t}{K_p}{\lambda _p}}}}}}}{{2M}} \\
&\times \sum\limits_{m = 1}^M {\sqrt {1 - v_m^2} } {F_{{{\widehat G}_{\max }}}}\left( { - \frac{{{P_t}{K_p}{\lambda _p}}}{{{\gamma _{{\rm{th}}}}}}\ln\kappa _m^{(1)}} \right).
\end{align}
Thus the low bound for the outage probability of the legacy transmission in the worst case can be computed as
\begin{align}\notag
\mathop {{\rm{lim}}}\limits_{{P_t} \to \infty } {\cal P}_{{\rm{out}},l}^{{\rm{wc}}} &\approx 1 - \frac{\pi }{{2M}}\sum\limits_{m = 1}^M {\sqrt {1 - v_m^2} } \\
&\times  {F_{{{\hat G}_{\max }}}}\left( { - \frac{{{P_t}{K_p}{\lambda _p}}}{{{\gamma _{{\rm{th}}}}}}\ln \kappa _m^{(1)}} \right).
\end{align}

\begin{figure}
  \centering
  \includegraphics[width=0.4\textwidth]{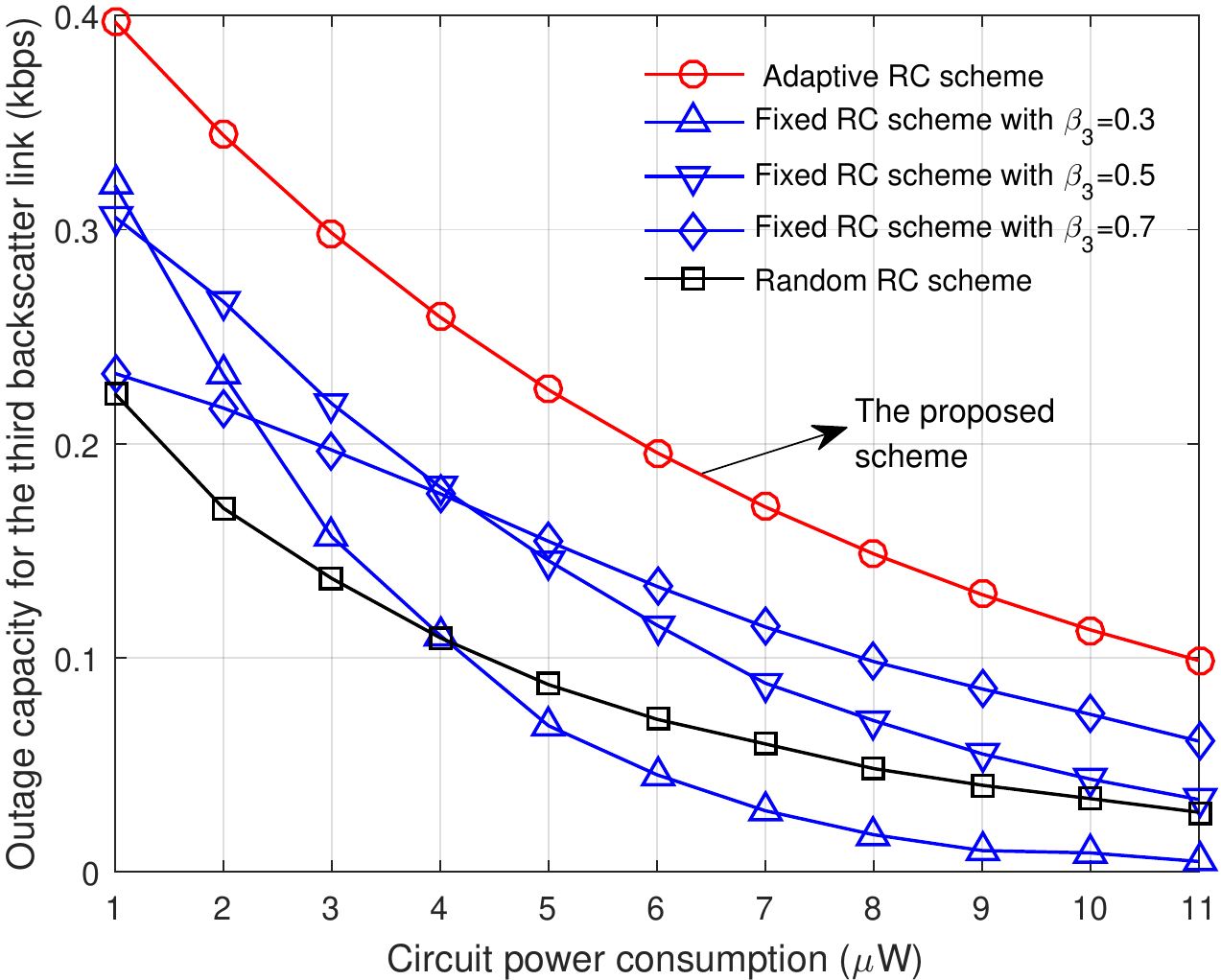}\\
  \caption{Outage capacity for the third backscatter link versus the circuit power consumption.}\label{fig2}
\end{figure}
{\color{black}For better understanding, we summarize  the  meaning of the derived outage probabilities as Table II.}

\begin{table}
\centering
\begin{tabular}{|l|l|}
\multicolumn{2}{c}{\textbf{{\color{black}Table II Meaning of the Derived Outage Probabilities}}}\\
\hline
\color{black} Notation  & \color{black} Meaning  \\
\hline
\color{black} $\mathcal{P}^b_{{\rm{out}},k}$  &  \color{black} Outage probability for the $k$-th backscatter link\\
\hline
\color{black} $\mathcal{P}_{{\rm{out}},b}^{\rm{bc}}$  & \color{black} The lowest outage probability for the backscatter system \\
\hline
\color{black} $\mathcal{P}_{{\rm{out}},b}^{\rm{wc}}$ &\color{black} The highest outage probability for the backscatter system\\
\hline
\color{black} $\mathcal{P}^l_{{\rm{out}},k}$ &  \color{black} Outage probability for the legacy  link when the $k$-th \\
                                             &     \color{black} backscatter link is selected\\
\hline
\color{black} $ {\cal P}_{{\rm{out}},l}^{{\rm{bc}}}$ & \color{black} The lowest outage probability for the legacy system \\
\hline
\color{black} ${\cal P}_{{\rm{out}},l}^{{\rm{wc}}}$ &\color{black} The highest outage probability for the legacy system\\
\hline
\end{tabular}
\end{table}

\section{Simulation and Discussion}
In this section, we evaluate the outage performance of AmBackComs.
{\color{black}In what follows, based on \cite{7981380,8327597,7996410,8170328}, we set   the basic parameter values as follows:   $K=3$, $W=1$ MHz, $T=1$ second, $P_{c,1}=P_{c,2}=P_{c,3}=8.9\;\mu$W,
$U=1$ kbps, and $\alpha=2.7$. The backscatter efficiency $\eta_k$ reflects  a loss of $1.1$ dB [15], and the noise power spectral density is set to be $-120$ dBm/Hz
 \cite{8170328}.}
For the legacy transmission, the given transmission rate is set as $10$ Mbps.
According to \cite{4494663}, the parameters of the considered non-linear EH model is set as
$E_{\max,1}=E_{\max,2}=E_{\max,3}=240$$\mu$W, $s_{11}=s_{12}=s_{13}=5000$ and $s_{21}=s_{22}=s_{23}=0.0002$.

Besides the above simulation parameters, the distances are set as: $d_p=10$ metres, $d_1^p=5$ metres, $d_2^p=3$ metres, $d_3^p=4$ metres, $d_{11}=4$ metres, $d_{12}=2$ metres, $d_{13}=3$ metres, $d_{21}=1.2$ metres, $d_{22}=2$ metres, $d_{23}=1.5$ metres, $d_1^s=7$ metres, $d_2^s=9$ metres, and $d_3^s=8$ metres.
{\color{black}According to \cite{8633928}, we suppose that the LT transmits its signal at $915$ MHz and the antenna gain for each BD is set to be $1.8$ dBi, while the antenna gains for the LT, the LR and the BRs are all set as $6$ dBi.}

\begin{figure}
  \centering
  \includegraphics[width=0.4\textwidth]{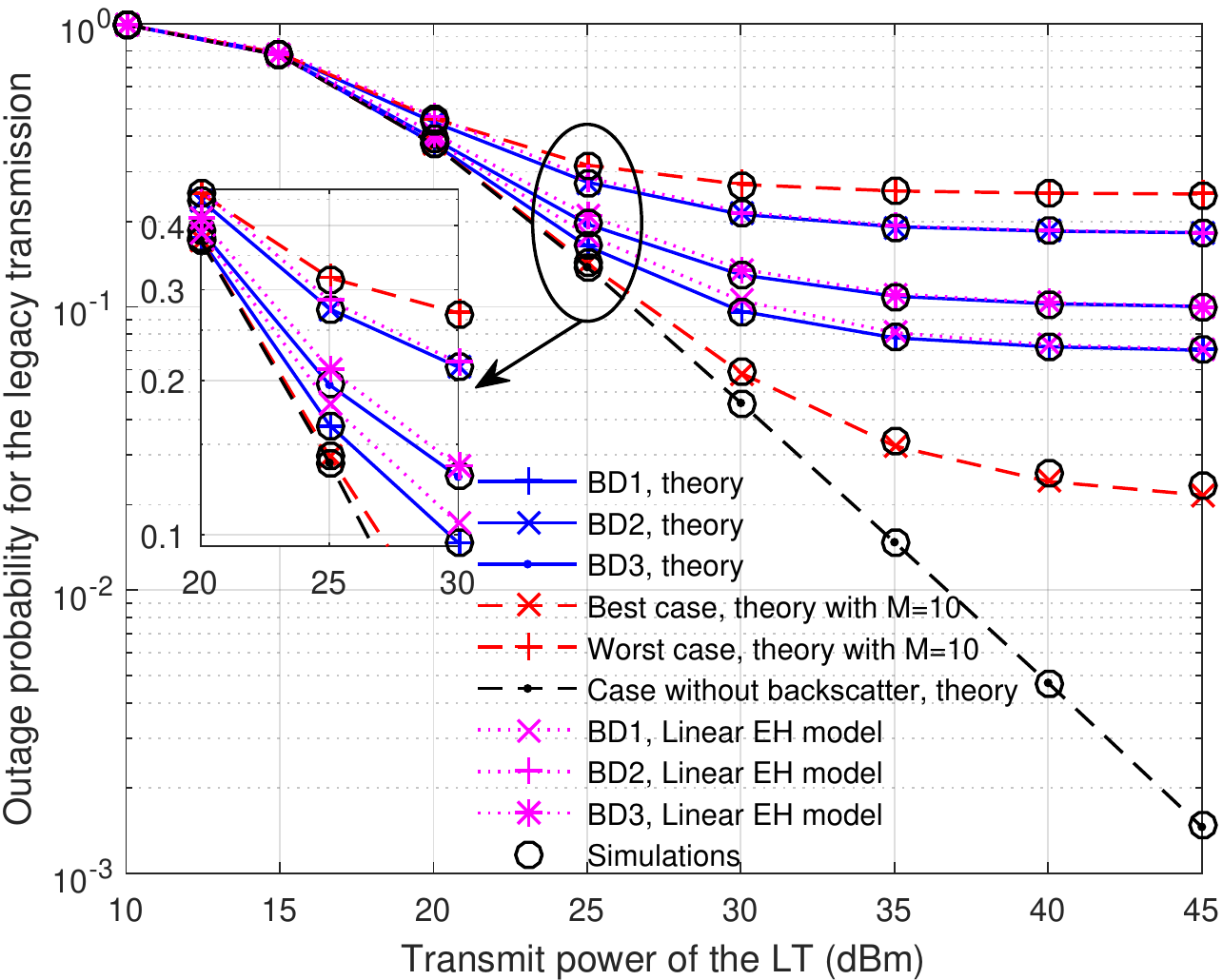}\\
  \caption{Outage probability of the legacy transmission versus the transmit power of the LT.}\label{fig4}
\end{figure}
\subsection{Outage performance analysis for the backscatter links}
Fig. 2 shows the outage probabilities of the backscatter system versus the transmit power of the LT $P_t$.
Note that the outage probability for the $k$-th ($k\in\{1,2,3\}$) backscatter link is computed based on \eqref{12} (\eqref{12-1} for the high SNR approximation)
while the outage probabilities under the best case and the worst case are determined by \eqref{13} and \eqref{14}, respectively. For the linear EH model, the conversion efficiency is fixed as $0.8$ throughout the simulations. 
As shown in this figure, we can see that our derived analytical results match well with simulation results that are obtained by over $1\times10^6$  Monte Carlo simulations  and marked by red circles. This observation demonstrates the correctness of our derived outage probabilities. It also shows that a small $M$, i.e., $M=10$, is sufficient to provide an accurate outage probability for Gaussian-Chebyshev approximation.
Besides, the outage probabilities under the linear EH model can not correctly characterize the outage performance of the backscatter system. For example, the outage performance under the linear model is over-estimated for the backscatter system.
In addition, we can also observe that all the outage probabilities decrease with the increasing of $P_t$ and when $P_t$ is large enough, all the outage probabilities get saturated. That is to say, there exists an  error floor caused by the interference from the legacy transmission.
Another observation is that the outage probability under the best case can achieve the best outage performance while the outage probability under the worst case is the highest as expected.
It is also found that choosing the backscatter link with the maximum SINR at the BR for backscattering can greatly reduce the  error floor and improve the outage performance of backscatter system.

Fig. 3 illustrates the outage capacity for the backscatter link versus the circuit power
consumption, where the third BD is selected to backscatter information and three schemes are employed at the BD, namely the proposed adaptive RC scheme, the fixed RC scheme and the random RC scheme.
Specifically, $P_t$ is set to be $20$~dBm. For the proposed scheme, the RC ratio is determined by \eqref{b1}. For the fixed RC scheme, the RC ratio is set to be $0.3$, $0.5$ and $0.7$, respectively. For the random RC scheme, the RC ratio follows a uniform distribution over the closed interval $[0,1]$.
It can be observed that as the circuit power consumption increases, the outage capacity decreases. This is because with the increasing of the circuit power consumption, the outage probability will increase due to the fact that most of received signals will be harvested to power the circuit operation, resulting in a smaller received SINR at the BR. Besides, a larger outage probability leads to a smaller outage capacity.
By comparisons, we can see that the proposed scheme can achieve the best performance since the proposed scheme provides more flexibility to utilize the resource efficiently.


\begin{figure}
  \centering
  \includegraphics[width=0.43\textwidth]{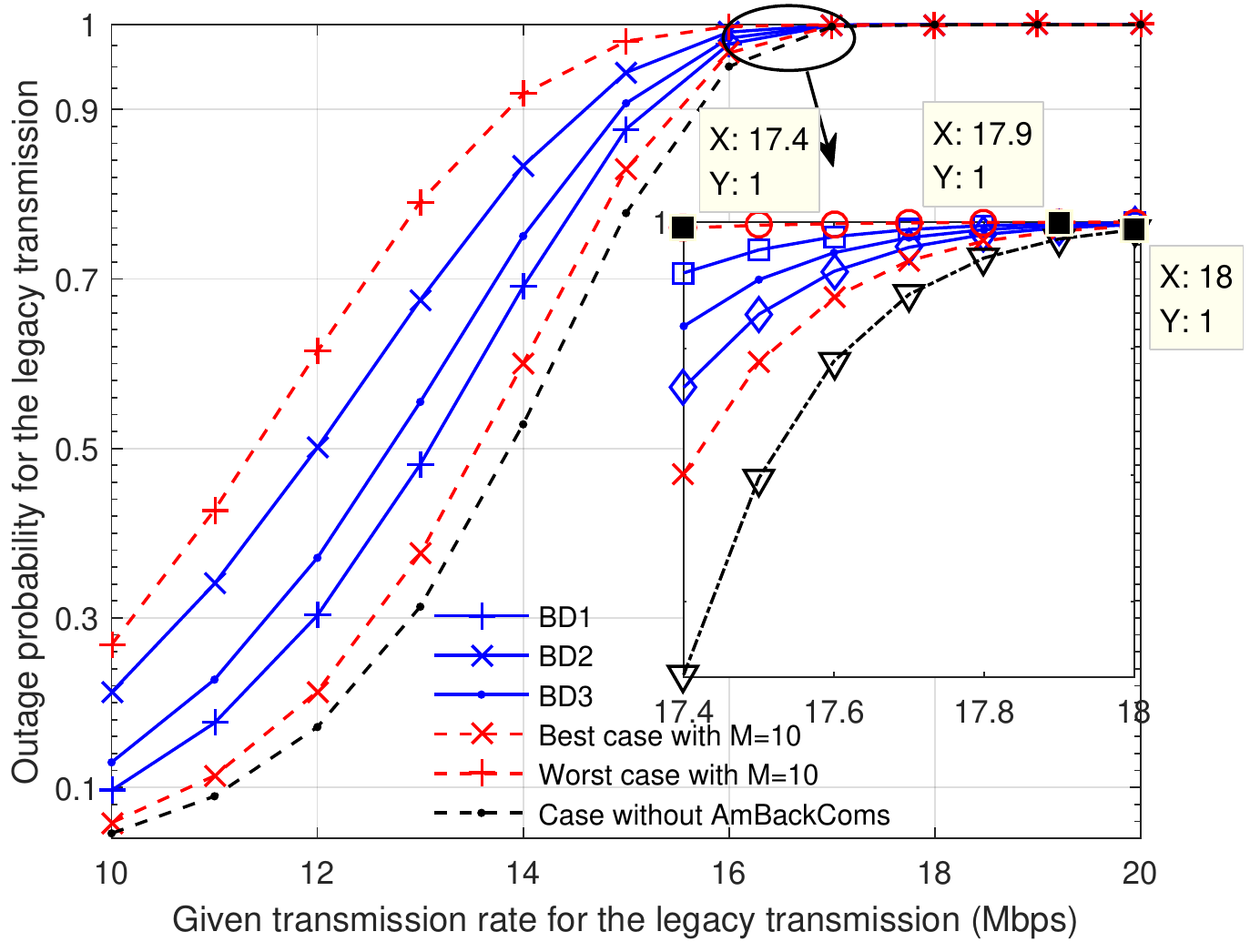}\\
  \caption{Outage probability of the legacy transmission versus the given transmission rate.}\label{fig5}
\end{figure}

\subsection{The impact of backscatter links on the legacy transmission}
Fig. 4 plots the outage probability for the legacy transmission as a function of the transmit power of the LT, where four cases are considered. The first case is the case without backscatter transmission. In this case, the legacy transmission will not be interfered and the outage probability for the legacy transmission is given by \eqref{19}. The second case is the case where the $k$-th ($k\in\{1,2,3\}$) BD is selected and the outage probability of the legacy transmission is determined by \eqref{15}. The third case is the best case where the backscatter link with the minimum interference  on the legacy link is selected for backscattering and the outage probability of this case is given by \eqref{20}. The fourth case is the worst case where the backscatter link with the maximum interference on the legacy link is selected and the outage probability is computed based on \eqref{25}. One observation is that there is an exact agreement between the derived results and the Monte Carlo simulation results.
\begin{figure*}[htbp]
\centering
\subfigure[]{
\begin{minipage}[t]{0.3\linewidth}
\centering
\includegraphics[width=2in,height=1.35in]{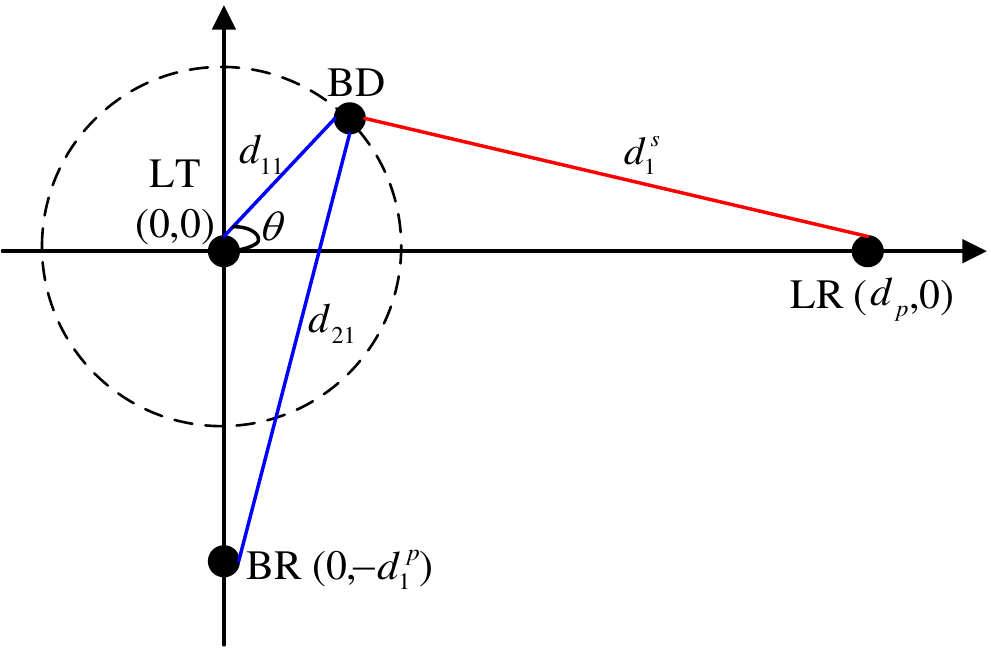}
\end{minipage}%
}%
\subfigure[]{
\begin{minipage}[t]{0.35\linewidth}
\centering
\includegraphics[width=2.55in,height=2in]{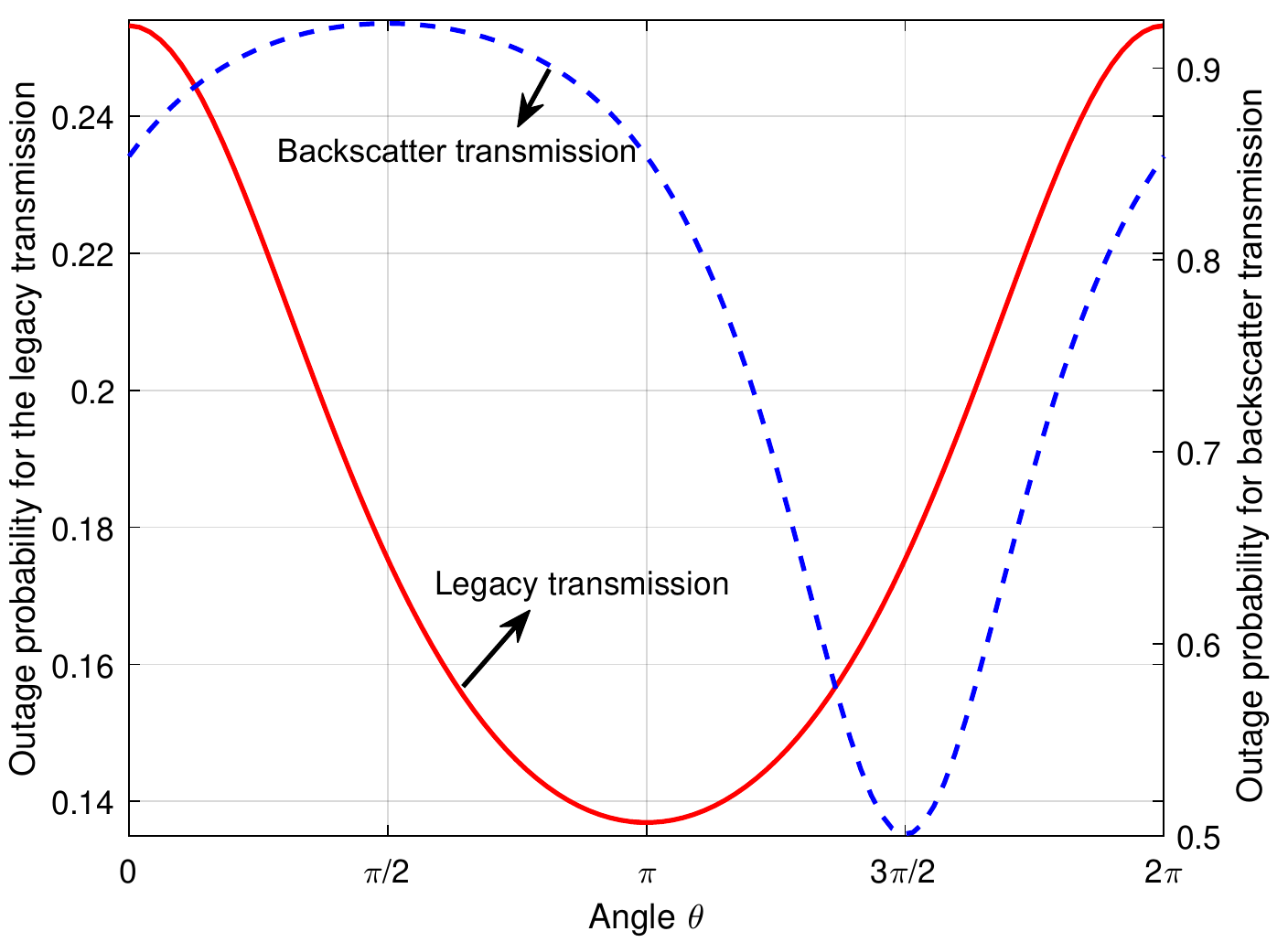}
\end{minipage}%
}%
\subfigure[]{
\begin{minipage}[t]{0.35\linewidth}
\centering
\includegraphics[width=2.55in,height=2in]{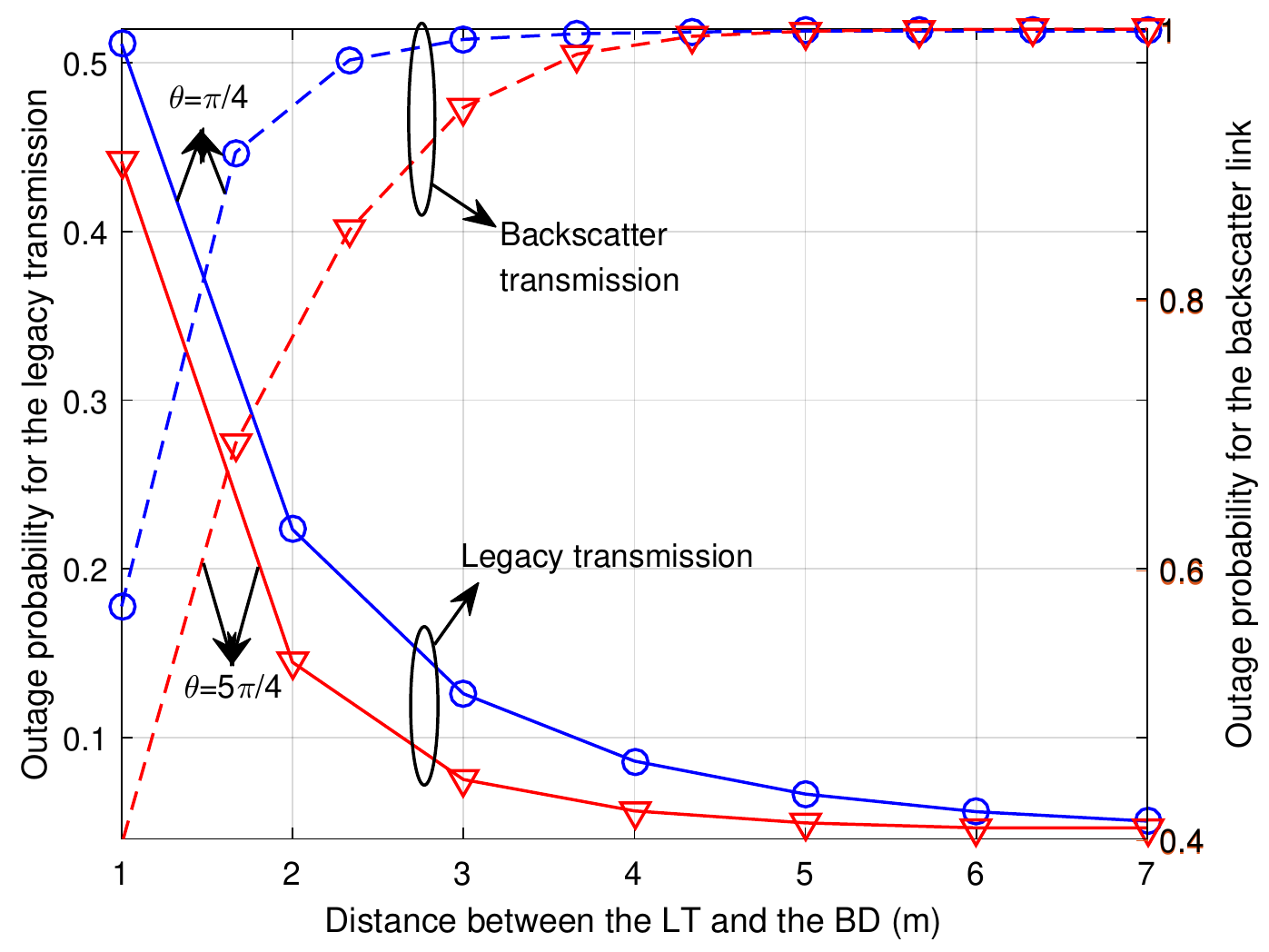}
\end{minipage}
}
\centering
\caption{(a) Relative locations among the LT, LR, BD and BR. (b) The impact of the location of the BD on outage probabilities for the legacy and backscatter transmissions. (c) Outage probabilities for the legacy and backscatter transmissions versus the distance between the LT and the BD.}
\end{figure*}
Besides, we also observe that the outage probability of the legacy transmission with an interference from the backscatter link
is generally inferior to that of the legacy transmission with no interference. Due to the existence of the interference, there exists an  error floor for the outage probability of the legacy transmission.
Another observation is that the outage probability under the best case is lowest among the cases with interferences while the outage probability under the worst case is highest. In addition,  the selected backscatter link with the minimum interference for backscattering is helpful to reduce  error floor and eliminate the bad influence as much as possible.
 Also, we can  find that the results with the linear model  can not catch the outage performance of the real systems, but the gap caused by the
inaccurate energy harvesting mode is smaller than the backscatter system.

Fig. 5 shows the outage probability of the legacy transmission versus the given
transmission rate $R_{\rm{th}}$, where $P_t$ is set as $30$~dBm. It can be observed that with the increasing of $R_{\rm{th}}$, all the outage probabilities firstly increase and then equal one. For example, when $R_{\rm{th}}$ is large enough, i.e., $R_{\rm{th}}\geq18$~Mbps, all the outage probabilities are always one.
This observation is refer to as rate ceiling. In addition, it can also be found that the outage probability of the best case is lowest and the corresponding rate ceiling is the largest.
\subsection{Trade-off for the outage performance between the legacy transmission and the backscatter link}
In order to study the impact of the location of the BD on the both legacy and backscatter links, we consider a specific scenario which consists of one LT, one LR, one BD and one BR, as shown in Fig. 6(a). In particular, the LT is located at the origin $(0,0)$. The LR and the BR lie on the x-axis with $(d_p,0)$ and the y-axis with $(0,-d_1^p)$, respectively. Assume that the BD moves on a circle of radius $d_{11}$ centered around the origin. Denote the angle between the radius and the x-axis as $\theta$. Then $d_{21}$ and $d^s_1$ are computed as ${d_{21}} = \sqrt {d_{11}^2 + {{\left( {d_1^p} \right)}^2} - 2{d_{11}}d_1^p\cos \left( {\theta  + \frac{\pi }{2}} \right)} $ and $d_1^s = \sqrt {d_{11}^2 + {{\left( {{d_p}} \right)}^2} - 2{d_{11}}{d_p}\cos \left( \theta  \right)} $, respectively.
Let $d_p$, $d_{11}$ and $d^p_1$ be $10$~m, $2$~m and $4$~m.

Fig. 6(b) plots the outage probabilities of the legacy transmission and the backscatter link versus $\theta$ for the considered scenario.
It can be observed that the outage probability of the legacy link decreases when the angle varies in the range of $[0,\pi]$ and increases within $[\pi,2\pi]$, while the outage probability of the backscatter transmission increases within the ranges of $[0,\frac{\pi}{2}]$ and $[\frac{3\pi}{2},2\pi]$ and decreases within $[\frac{\pi}{2},\frac{3\pi}{2}]$.
It can be found that the minimum outage probabilities for the legacy and the backscatter transmissions are achieved at $\theta=\pi$ and $\theta=\frac{3\pi}{2}$, respectively. With $\theta\in[0,\frac{\pi}{2}]$ (or $\theta\in[\pi,\frac{3\pi}{2}]$), a better outage performance for the legacy (or backscatter) link  is achieved at the cost of the backscatter (legacy) transmission's performance. With $\theta\in[\frac{\pi}{2},\pi]$, a win-win situation will be achieved and when $\theta\in[\frac{3\pi}{2},2\pi]$, the outage performance of both the legacy and the backscatter links will become worse, which needs to be avoided.
It is worth noting that the win-win area can be enlarged by adjusting the location of the BR with $d^p_1$ unchanged. For example, when the BR in Fig. 6(a) moves away from the LR in a clockwise direction with $d^p_1$ unchanged, the bule dotted line in Fig. 6(b) will move the same angle to the left and the win-win area also increases. When the BR is located at $(-d_1^p,0)$ (the opposite direction of the LR), the win-win area is $\theta\in[0,\pi]$, which is the largest. When the BR is located at $(d_1^p,0)$, there is no win-win area in this case.
Besides, we can also get some insights for the deployment of the BD. For example, we can choose a proper angle to minimize the outage probability of the backscatter link while satisfying a given outage constraint for the legacy transmission.

Fig. 6(c) shows the impact of $d_{11}$ on the outage probabilities of the legacy transmission and the backscatter link, where $\theta$ is fixed as $\frac{\pi}{4}$ and $\frac{5\pi}{4}$, respectively.
It can be observed that with the increasing of $d_{11}$, the outage probabilities of the legacy transmission decrease while the outage probabilities of the backscatter link increase. This is because the received RF power decreases with the increasing of $d_{11}$, resulting in a smaller backscattered signal at both the LR and the BR. It can also be found that under the given outage constraint for the legacy link, it is better to choose a smaller $d_{11}$ to achieve a better outage performance of the backscatter transmission.

\section{Conclusions}
In this work, we have studied the  outage performance for AmBackComs. 
Specifically,  we have proposed an adaptive RC to minimize the outage probability  for any given backscatter link.
 With a given backscatter link and the  optimal RC ratio, we  derived  the outage probabilities for the backscatter system and the legacy transmission, respectively.
Besides, the best and worst outage performances of the backscatter system and the legacy transmission have been investigated, respectively.
Simulation results have verified the correctness of our analytical results and provided practical insights into the
impacts of the co-channel interference,   the EH model, and the location of BDs. {\color{black}Key observations have been summarized as follows.
Firstly, the co-channel
interference leads to the outage saturation phenomenon
in the  backscatter  and  legacy links. Besides, the backscatter transmission leads to a rate ceiling for the legacy link. Secondly, selecting the  backscatter link with the lowest interference on the LR ensures that the performance loss of the legacy link  is kept to a minimum. Thirdly,  the conventionally used linear EH model will result in an over-estimated outage performance for the backscatter link while the impact on the legacy link is very small.}

\section*{Appendix A}
We first compute $\mathbb{P}({S_k} \le x,|{h_{1k}}{|^2} \!\!\geq\! \frac{{{b_k}}}{{{a_k}}})$ as
\begin{align}\label{A1}\notag
&\mathbb{P}\!({S_k} \le x,|{h_{1k}}{|^2} \geq \frac{{{b_k}}}{{{a_k}}})\\ \notag
&\overset{\text{(a)}}{=}\int_{\frac{{{b_k}}}{{{a_k}}}}^{ + \infty } {\!\left[\! {1- \exp \left( \!{ - \frac{x}{{{\lambda _{2k}}\!\left(\! {{a_k}y \!-\! {b_k}} \!\right)\!}}} \!\right)}\! \right]\!\!\frac{{e^ { \!{ - \frac{y}{{{\lambda _{1k}}}}} \!}\!}}{{{\lambda _{1k}}}}dy}\\ \notag
&\overset{\text{(b)}}{=}\left(\! {1 - \frac{1}{{{\lambda _{1k}}{a_k}}}\!\!\int_0^{ + \infty } {\exp\! \left(\! { - \frac{x}{{{\lambda _{2k}}z}} - \frac{z}{{{a_k}{\lambda _{1k}}}}} \!\right)\!} } dz\right)\\
&\times \exp\left(-\frac{b_k}{a_k\lambda_{1k}}\right),\tag{A.1}
\end{align}
where steps (a) and (b) follows from $y=|{h_{1k}}{|^2}$ and $z={{a_k}y - {b_k}}$, respectively.
Then $F_{S_k}(x)$ can be calculated as
\begin{align}\label{A2}\notag
\!\!\!\!F_{S_k}(x)&=\mathbb{P}({S_k} \le x\big||{h_{1k}}{|^2} \ge \frac{{{b_k}}}{{{a_k}}})\\ \notag
&=\frac{{ \mathbb{P}({S_k} \le x,|{h_{1k}}{|^2} \ge \frac{{{b_k}}}{{{a_k}}})}}{{ \mathbb{P}(|{h_{1k}}{|^2} \ge \frac{{{b_k}}}{{{a_k}}})}}\\ \notag
&={1 - \frac{1}{{{\lambda _{1k}}{a_k}}}\!\!\int_0^{ + \infty } \!\!\!\!\!\!\!{\exp\! \left(\! { - \frac{x}{{{\lambda _{2k}}z}} - \frac{z}{{{a_k}{\lambda _{1k}}}}} \!\right)} }dz\\
&\overset{\text{(c)}}{=}1 - \frac{1}{{{\lambda _{1k}}{a_k}}}\sqrt {\frac{{4{a_k}{\lambda _{1k}}x}}{{{\lambda _{2k}}}}} {K_1}\left( {\sqrt {\frac{{4x}}{{{\lambda _{2k}}{\lambda _{1k}}{a_k}}}} } \right),\tag{A.2}
\end{align}
where step (c) holds based on [3.324] in \cite{b1} and $K_1(\cdot)$ is the modified Bessel function of the second kind.

Further, the PDF of $S_k$ conditioned on $|{h_{1k}}{|^2} \!\!\geq\! \frac{{{b_k}}}{{{a_k}}}$ is
\begin{align}\label{A3}\notag
f_{S_k}(x)&=\frac{{\partial \mathbb{P}({S_k} \le x\big||{h_{1k}}{|^2} \ge \frac{{{b_k}}}{{{a_k}}})}}{{\partial x}}\\ \notag
&=\frac{1}{{{\lambda _{1k}}{\lambda _{2k}}{a_k}}}\int_0^{ + \infty }  \frac{1}{z}\exp \left( { - \frac{x}{{{\lambda _{2k}}z}} - \frac{z}{{{a_k}{\lambda _{1k}}}}} \right)dz\\
&\overset{\text{(d)}}{=}\frac{2}{{{\lambda _{1k}}{\lambda _{2k}}{a_k}}}{K_0}\left( {2\sqrt {\frac{x}{{{\lambda _{1k}}{\lambda _{2k}}{a_k}}}} } \right),\tag{A.3}
\end{align}
where step (d) follows by [3.478] in \cite{b1} and $K_0(\cdot)$ is the modified Bessel function of the second kind.

\section*{Appendix B}
Following $|{h_{1k}}{|^2}\sim \exp(\frac{1}{\lambda_{1k}})$, we can compute $I_1$ as
\begin{align}\label{9}
I_1=1-\exp\left(-\frac{b_k}{a_k\lambda_{1k}}\right).\tag{B.1}
\end{align}
Then the main difficulty is how to compute $I_2$.
In order to obtain $I_2$, we first provide the following proposition to achieve both the CDF and PDF with respect to $S_k$ conditioned on $|{h_{1k}}{|^2} \geq \frac{{{b_k}}}{{{a_k}}}$.

By means of Proposition 1, we can calculate $I_2$ as
\begin{align}\label{11}\notag
I_2&=\mathbb{P}\left(\!\frac{S_k}{{{P_t}{K_{k}^p}{{\left| {g_k^p} \right|}^2} + {\sigma ^2}}}< \gamma^b_{\rm{th}}\bigg| |{h_{1k}}{|^2} \geq \frac{{{b_k}}}{{{a_k}}}\right)\\ \notag
&\times \mathbb{P}\left(|{h_{1k}}{|^2} \geq \frac{{{b_k}}}{{{a_k}}}\right)\\ \notag
&\overset{\text{(a)}}{=}\left( {1 - {I_1}} \right)\bigg[ \underbrace{\int_{\gamma _{{\rm{th}}}^b{\sigma ^2}}^{ + \infty }  \exp \left( { - \frac{x}{{\gamma _{{\rm{th}}}^b{P_t}{K_{k}^p}{\lambda _{k}^p}}}} \right){f_{{S_k}}}\left( {x} \right)dx}_{I_{2-1}} \\
&\times\exp \left( {\frac{{{\sigma ^2}}}{{{P_t}{K_{k}^p}{\lambda _{k}^p}}}} \right)+ {F_{{S_k}}}\left( {\gamma _{{\rm{th}}}^b{\sigma ^2}} \right) \bigg], \tag{B.2}
\end{align}
where step (a) follows by letting $|{g_k^p}{|^2}\sim \exp(\frac{1}{\lambda_{k}^p})$ and considering two cases, i.e.,  $S_k>\gamma _{{\rm{th}}}^b{\sigma ^2}$  and $S_k\leq\gamma _{{\rm{th}}}^b{\sigma ^2}$.

Further, $I_{2-1}$ can be rewritten as
\begin{align}\label{11-1}\notag
I_{2-1}
&=I_{2-2}-\int_{0}^{ \gamma _{{\rm{th}}}^b{\sigma ^2}}  \exp \left( { - \frac{x}{{\gamma _{{\rm{th}}}^b{P_t}{K_{k}^p}{\lambda _{k}^p}}}} \right){f_{{S_k}}}\left( {x} \right)dx\\
&\overset{\text{(b)}}{=}\Theta_k-\int_{0}^{ \gamma _{{\rm{th}}}^b{\sigma ^2}}  \exp \left( { - \frac{x}{{\gamma _{{\rm{th}}}^b{P_t}{K_{k}^p}{\lambda _{k}^p}}}} \right){f_{{S_k}}}\left( {x} \right)dx, \tag{B.3}
\end{align}
where $I_{2-2}=\int_{0}^{ + \infty }  \exp \left( { - \frac{x}{{\gamma _{{\rm{th}}}^b{P_t}{K_{k}^p}{\lambda _{k}^p}}}} \right){f_{{S_k}}}\left( {x} \right)dx$, step (b) follows from Appendix C and $\Theta_k=- {\vartheta _k}\gamma _{{\rm{th}}}^b{\rm{Ei}}( - {\vartheta _k}\gamma _{{\rm{th}}}^b)\exp \left( {{\vartheta _k}\gamma _{{\rm{th}}}^b} \right)$ with ${\vartheta _k}=\frac{{{P_t}{K_{k}^p}{\lambda _{k}^p}}}{{{\lambda _{1k}}{\lambda _{2k}}{a_k}}}$ and the exponential integral function ${\rm{Ei}}(\varrho)=\int_{-\infty}^{\varrho}t^{-1}e^{t}dt$.

\section*{Appendix C}
Let $\alpha_1$ and $\alpha_2$ denote $\frac{1}{{\gamma _{{\rm{th}}}^b{P_t}K_k^p\lambda _k^p}}$ and $\frac{1}{{{\lambda _{1k}}{\lambda _{2k}}{a_k}}}$, respectively. Then $I_{2-2}$ can be calculated as
\begin{align}\label{B1}\notag
I_{2-2}&=\frac{2}{{{\lambda _{1k}}{\lambda _{2k}}{a_k}}}\int_0^{{\rm{ + }}\infty } {{e^{ - {\alpha _1}x}}{K_0}\left( {2\sqrt {{\alpha _2}x} } \right)dx}\\ \notag
&\overset{\text{(a)}}{=}\frac{{{e^{\frac{{{\alpha _2}}}{{2{\alpha _1}}}}}}}{{{\lambda _{1k}}{\lambda _{2k}}{a_k}\sqrt {{\alpha _1}{\alpha _2}} }}{\left( {\Gamma (1)} \right)^2}{W_{ - \frac{1}{2},0}}\left( {\frac{{{\alpha _2}}}{{{\alpha _1}}}} \right)\\ \notag
&=\frac{1}{{{\lambda _{1k}}{\lambda _{2k}}{a_k}{\alpha _1}}}\int_0^{ + \infty } {\frac{{{e^{ - \frac{{{\alpha _2}t}}{{{\alpha _1}}}}}}}{{1 + t}}dt} \\ \notag
&=\frac{{ - {e^{\frac{{{\alpha _2}}}{{{\alpha _1}}}}}}}{{{\lambda _{1k}}{\lambda _{2k}}{a_k}{\alpha _1}}}{\rm{Ei}}\left( { - \frac{{{\alpha _2}}}{{{\alpha _1}}}} \right)\\
&\overset{\text{(b)}}{=}- {\vartheta _k}\gamma _{{\rm{th}}}^b{\rm{Ei}}( - {\vartheta _k}\gamma _{{\rm{th}}}^b)\exp \left( {{\vartheta _k}\gamma _{{\rm{th}}}^b} \right), \tag{C.1}
\end{align}
where step (a) holds based on [6.614] in \cite{b1}; $\Gamma(x)=\int_0^{{\rm{ + }}\infty } {{t^{x - 1}}{e^{ - t}}dt} $ is a gamma function;
${W_{\mu ,\nu }}\left( x \right)$ is a Whittaker function which is given by ${W_{\mu ,\nu }}\left( x \right) = \frac{{{x^{\nu  + 0.5}}{e^{ - \frac{x}{2}}}}}{{\Gamma \left( {\nu  - \mu  + 0.5} \right)}}\int_0^{ + \infty } {{e^{ - xt}}{t^{\nu  - \mu  - 0.5}}{{\left( {1 + t} \right)}^{\nu  + \mu  - 0.5}}dt} $; ${\rm{Ei}}(\varrho)=\int_{-\infty}^{\varrho}t^{-1}e^{t}dt$ denotes the exponential integral function; step (b) follows by letting ${\vartheta _k}=\frac{{{P_t}{K_{k}^p}{\lambda _{k}^p}}}{{{\lambda _{1k}}{\lambda _{2k}}{a_k}}}$.

\section*{Appendix D}
In \eqref{a10}, the term ${\int_0^{\gamma _{{\rm{th}}}^b{\sigma ^2}} {\exp } \left( { - \frac{x}{{\gamma _{{\rm{th}}}^b{P_t}K_k^p\lambda _k^p}}} \right){f_{{S_k}}}\left( x \right)dx}$ is not a closed-form expression. Thus the purpose of this appendix to approximate this term based on a high transmit power of the LT.
\begin{align}\notag
&{\int_0^{\gamma _{{\rm{th}}}^b{\sigma ^2}} {\exp } \left( { - \frac{x}{{\gamma _{{\rm{th}}}^b{P_t}K_k^p\lambda _k^p}}} \right){f_{{S_k}}}\left( x \right)dx}\\ \notag
&\mathop {\rm{ = }}\limits^{(a)} \frac{{ - 2}}{{{\lambda _{1k}}{\lambda _{2k}}{a_k}}}\! \int_0^{\gamma _{{\rm{th}}}^b{\sigma ^2}} \!\!\!\!\!\!\!{\exp } \left( \!\!{ - \frac{x}{{\gamma _{{\rm{th}}}^b{P_t}K_k^p\lambda _k^p}}}\! \right)\!\ln\! \left(\! {\sqrt {\frac{x}{{{\lambda _{1k}}{\lambda _{2k}}{a_k}}}} } \right)\!dx \\ \notag
&- \frac{2}{{{\lambda _{1k}}{\lambda _{2k}}{a_k}}}\int_0^{\gamma _{{\rm{th}}}^b{\sigma ^2}} {c_0 \exp \left( { - \frac{x}{{\gamma _{{\rm{th}}}^b{P_t}K_k^p\lambda _k^p}}} \right)dx}\\ \notag
&\mathop {\rm{ = }}\limits^{(b)} \frac{{ - 4}}{{{\lambda _{1k}}{\lambda _{2k}}{a_k}}}\!\!\int_0^{\sqrt{\gamma _{{\rm{th}}}^b}{\sigma }} \!\!\!\!\!\!{y\exp } \left(\!\! { - \frac{{{y^2}}}{{\gamma _{{\rm{th}}}^b{P_t}K_k^p\lambda _k^p}}} \!\right)\!\ln\!\! \left(\! {\frac{y}{{\sqrt {{\lambda _{1k}}{\lambda _{2k}}{a_k}} }}} \right)\!dy \\
&- 2\gamma _{{\rm{th}}}^b{\vartheta _k}c_0\left( {1 - \exp \left( { - \frac{{{\sigma ^2}}}{{{P_t}K_k^p\lambda _k^p}}} \right)} \right). \tag{D.1}
\end{align}
where step (a) holds from ${K_0}\left( x \right) \approx    - \ln \left( {\frac{x}{2}} \right) - c_0$ at $x \to 0$ [eq. (8.446), \cite{b1}] and the fact that  the noise power is very small in practical communications, i.e., $\sigma^2 \to 0$; $c_0 \approx 0.5772$ is the Euler constant; step (b) is derived from the variable substitution, i.e., $y=\sqrt{x}$.

Based on  the   fourier series of $\exp \left( { - \frac{{{x^2}}}{{\gamma _{{\rm{th}}}^b{P_t}K_k^p\lambda _k^p}}} \right)$ at $x \to 0$ [eq.(1.211.3), \cite{b1}], we have the following approximation when the transmit power of the LT is high, given by
\begin{align}\notag
&\int_0^{\sqrt {\gamma _{{\rm{th}}}^b} \sigma } {y\exp } \left( { - \frac{{{y^2}}}{{\gamma _{{\rm{th}}}^b{P_t}K_k^p\lambda _k^p}}} \right)\ln \left( {\frac{y}{{\sqrt {{\lambda _{1k}}{\lambda _{2k}}{a_k}} }}} \right)dy\\ \notag
 &\approx \int_0^{\sqrt {\gamma _{{\rm{th}}}^b} \sigma } y \ln \left( {\frac{y}{{\sqrt {{\lambda _{1k}}{\lambda _{2k}}{a_k}} }}} \right)dy\\ \notag
  &- \int_0^{\sqrt {\gamma _{{\rm{th}}}^b} \sigma } {\frac{{{y^3}}}{{\gamma _{{\rm{th}}}^b{P_t}K_k^p\lambda _k^p}}} \ln \left( {\frac{y}{{\sqrt {{\lambda _{1k}}{\lambda _{2k}}{a_k}} }}} \right)dy \\ \notag
 &=\frac{1}{2}\gamma _{{\rm{th}}}^b{\sigma ^2}\left( {\ln \left( {\frac{{\sqrt {\gamma _{{\rm{th}}}^b} \sigma }}{{\sqrt {{\lambda _{1k}}{\lambda _{2k}}{a_k}} }}} \right) - \frac{1}{2}} \right)\\
& - \frac{{{{{\gamma _{{\rm{th}}}^b} }}{\sigma ^4}}}{{4{P_t}K_k^p\lambda _k^p}}\left( {\ln \left( {\frac{{\sqrt {\gamma _{{\rm{th}}}^b} \sigma }}{{\sqrt {{\lambda _{1k}}{\lambda _{2k}}{a_k}} }}} \right) - \frac{1}{4}} \right).\tag{D.2}
\end{align}

\section*{Appendix E}
We first provide the CDF of $G_k$. Similar to \eqref{A1}, $\mathbb{P}\!\left(\!G_k\leq x, |{h_{1k}}{|^2} \!> \! \frac{{{b_k}}}{{{a_k}}}\!\right)$ is given by
\begin{align}\label{C1}\notag
&\mathbb{P}\left(\!G_k\leq x, |{h_{1k}}{|^2} >  \frac{{{b_k}}}{{{a_k}}}\!\right)\\ \notag
&=\left( {1 \!-\! \frac{1}{{{\lambda _{1k}}{a_{ik}}}}\!\sqrt {\frac{{4{a_{ik}}{\lambda _{1k}}x}}{{{\lambda _{k}^s}}}} \!{K_1}\left(\! {\sqrt {\frac{{4x}}{{{\lambda _{1k}}{\lambda _{k}^s}{a_{ik}}}}} }\right)} \right) \\
&\times \exp\left(-\frac{b_k}{a_k\lambda_{1k}}\right).\tag{E.1}
\end{align}
Then the CDF of $\min \left( {{G_1},{G_2}, \ldots ,{G_K}} \right)$ conditioned on $\bigcap\limits_{k = 1}^K {|{h_{1k}}{|^2} > \frac{{{b_k}}}{{{a_k}}}}$, denoted by $F_{G_{\min}}(x)$, is given by
\begin{align}\label{C2}\notag
&F_{G_{\min}}(x) \\ \notag
&=\mathbb{P}\left(\!\min \left( {{G_1},{G_2}, \ldots ,{G_K}} \right)\!\leq x\bigg|\bigcap\limits_{k = 1}^K {|{h_{1k}}{|^2} > \frac{{{b_k}}}{{{a_k}}}}\!\right)\!\\ \notag
&=1-\mathbb{P}\left( {{G_1} > x,{G_2} > x, \ldots ,{G_K} > x}\bigg|\bigcap\limits_{k = 1}^K {|{h_{1k}}{|^2} > \frac{{{b_k}}}{{{a_k}}}} \right)\\ \notag
&=1-\prod\limits_{k = 1}^K {\frac{\mathbb{P}{\left( {{G_k} > x,|{h_{1k}}{|^2} > \frac{{{b_k}}}{{{a_k}}}} \right)\prod\limits_{i = 1,i \ne k}^K \mathbb{P}{\left( {|{h_{1i}}{|^2} > \frac{{{b_i}}}{{{a_i}}}} \right)} }}{{{\mathcal{P}_1}}}} \\
&=1-\prod\limits_{k = 1}^K  \left[ {\frac{{1}}{{{\lambda _{1k}}{a_{ik}}}}\sqrt {\frac{{4{a_{ik}}{\lambda _{1k}}x}}{{\lambda _k^s}}} {K_1}\left( {\sqrt {\frac{{4x}}{{{\lambda _{1k}}\lambda _k^s{a_{ik}}}}} } \right)} \right].\tag{E.2}
\end{align}

\section*{Appendix F}
Based on the expression of $\widehat{G}_k$, we first determine the CDF of $\widehat{G}_k$ as
\begin{align}\label{E.1}\notag
&\mathbb{P}\left(\widehat{G}_k\leq x\right)\\ \notag
&=\mathbb{P}\left(G_k\leq x,|{h_{1k}}{|^2} \!> \! \frac{{{b_k}}}{{{a_k}}}\right)+\mathbb{P}\left(|{h_{1k}}{|^2} \!\leq \! \frac{{{b_k}}}{{{a_k}}}\right)\\ \notag
&=1 - \frac{1}{{{\lambda _{1k}}{a_{ik}}}}\sqrt {\frac{{4{a_{ik}}{\lambda _{1k}}x}}{{\lambda _k^s}}} {K_1}\left( {\sqrt {\frac{{4x}}{{{\lambda _{1k}}\lambda _k^s{a_{ik}}}}} } \right) \\
&\times \exp \left( { - \frac{{{b_k}}}{{{a_k}{\lambda _{1k}}}}} \right).\tag{F.1}
\end{align}
Then the CDF of $\widehat{G}_{\max}$ can be calculated as
\begin{align}\label{E.2}\notag
&F_{\widehat{G}_{\max}}(x)\\ \notag
&=\mathbb{P}\left(\max \left( {{\widehat{G}_1},{\widehat{G}_2}, \ldots ,{\widehat{G}_K}} \right)\!\leq x\right)\\ \notag
&=\prod\limits_{k = 1}^K \left( {1 - \frac{1}{{{\lambda _{1k}}{a_{ik}}}}\sqrt {\frac{{4{a_{ik}}{\lambda _{1k}}x}}{{\lambda _k^s}}} {K_1}\left( {\sqrt {\frac{{4x}}{{{\lambda _{1k}}\lambda _k^s{a_{ik}}}}} } \right)} \right)\\
&\times \exp \left( { - \frac{{{b_k}}}{{{a_k}{\lambda _{1k}}}}} \right) + 1 - \exp \left( { - \frac{{{b_k}}}{{{a_k}{\lambda _{1k}}}}} \right). \tag{F.2}
\end{align}

\ifCLASSOPTIONcaptionsoff
  \newpage
\fi
\bibliographystyle{IEEEtran}
\bibliography{refa}

\end{document}